\documentclass[twocolumn,a4paper, nofootinbib]{revtex4-1}

\usepackage[T1]{fontenc}
\usepackage{amsmath}
\usepackage{amsthm}
\usepackage{amssymb}
\usepackage{amsfonts}
\usepackage{layout}
\usepackage{soul}
\usepackage{graphicx}
\usepackage{verbatim}
\usepackage{dsfont}
\usepackage{color}

\newcommand{\ket}[1]{\left| #1 \right>} 
\newcommand{\bra}[1]{\left< #1 \right|} 
\newcommand{\braket}[2]{\left< #1 \vphantom{#2} \right|
 \left. #2 \vphantom{#1} \right>} 

\newcommand{\conj}[1]{\overline{#1} }

 \def\bC {\mathbb{C}}
  \def\bZ {\mathbb{Z}}

\def\cA{{\cal{A}}}

\def\cR{\cal{R}}

\def\eye{\mathds{1}}

\def\hatH{{\hat{H}}}

\def\hatO{{\hat{O}}}

\def\hatsigma{{\hat{\sigma}}}

\newcommand{\norm}[1]{\left\lVert#1\right\rVert}

\DeclareMathOperator*{\argmin}{arg\,min}

\begin{document}
\title{Time Evolution and Deterministic Optimisation of Correlator Product States}

\author{Vid Stojevic$^{1}$, Philip Crowley$^{1}$,  Tanja \DJ uri\'c $^{2}$, Callum Grey $^{1}$ Andrew Green$^{1}$ }
\affiliation{
$^1$ London Centre for Nanotechnology, 17-19 Gordon St, London, WC1H 0AH \\
$^2$ Instytut Fizyki im. M. Smoluchowskiego, Uniwersytet Jagiello\'nski, \L{}ojasiewicza 11, 30-348 Krak\'ow, Poland
}

\begin{small}
\begin{abstract}
We study a restricted class of correlator product states (CPS) for a spin-half chain in which each spin is contained in just two overlapping plaquettes. This class is also a restriction upon matrix product states (MPS) with local dimension $2^n$ ($n$ being the size of the overlapping regions of plaquettes) equal to the bond dimension. We investigate the trade-off between gains in efficiency due to this restriction against losses in fidelity. The time-dependent variational principle formulated for these states is numerically very stable. Moreover, it shows significant gains in efficiency compared to the naively related matrix product states - the evolution or optimisation scales as $2^{3n}$ for the correlator product states versus $2^{4n}$ for the unrestricted matrix product state. However, much of this advantage is offset by a significant reduction in fidelity. Correlator product states break the local Hilbert space symmetry by the explicit selection of a local basis. We investigate this dependence in detail and formulate the broad principles under which correlator product states may be a useful tool. In particular, we find that scaling with overlap/bond order may be more stable with correlator product  states allowing a more efficient extraction of critical exponents - we present an example in which the use of correlator product states is several orders of magnitude quicker than matrix product states. 

\end{abstract}

\maketitle

\tableofcontents

\vspace{0.2in}

Variational Ansaetze have underpinned many advances in the study of correlated quantum systems, BCS superconductivity and the Laughlin wavefunctions being notable examples. The essence is to identify key physical features of a quantum state such that they can be parametrised by relatively few numbers, providing an analytically or numerically tractable approximation. The idea of covering the system with coupled plaquette clusters is basis of several ansaetze. Examples include coupled cluster methods \cite{darradi2008ground}, contractor renormalization methods \cite{altman2002plaquette} and various hierarchical mean-field approaches \cite{isaev2009hierarchical}. Within all of these methods, the configurations and correlations of the single plaquette or cluster are evaluated exactly, however the inter-plaquette couplings are essentially treated in mean-field. Correlator product states (CPS - also referred to as entangled plaquette states) \cite{2009PhRvB..80x5116C, 2009NJPh...11h3026M, 2010NJPh...12j3039M, 2010NJPh...12j3008M,1367-2630-11-8-083026,1367-2630-12-10-103039,al2011capturing,neuscamman2012correlator,djuric2014interaction,djuric2016pfaffian} go beyond this mean-field by fully parametrising the wavefunction on a set of overlapping plaquettes. Consistency between the wavefunctions in the regions of overlap communicates entanglement across the system. 

The realisation of the central importance of the entanglement structure of quantum many-body states has led to the introduction of many variational families of tensor network states that efficiently parametrise it.  The progenitor of these are matrix product states (MPS) \cite{Fannes:1992uq,verstraete2008matrix}.
MPS have been extended to higher dimensions using projected entangled pair states (PEPS) \cite{2004cond.mat..7066V},  and to accommodate various aspects of the renormalisation of entanglement at critical points \cite{2007PhRvL..99v0405V,2008PhRvL.101k0501V,2011arXiv1102.5524H}. Such states naturally obey an 'area law', which, in any dimension, refers to the property that the entanglement entropy of a large enough region scales not with the volume, but with the area of the boundary of that region. It is proven for gapped systems in one dimension \cite{1742-5468-2007-08-P08024, gottesman2010entanglement} - and suspected in higher dimensions - that  groundstates obey an area law and so can be efficiently described by tensor networks with relatively few components.  

However, the ability to represent a state efficiently in no way implies that quantities such as expectation values of local operators can be calculated efficiently. Indeed, it has been demonstrated \cite{2007PhRvL..98n0506S} that, unlike the computational complexity of contracting an arbitrary MPS, which is in P, the computational complexity of contracting a general PEPS network - the natural higher dimensional analogue -  is $\#$P-complete. In practice, except for very special tensor networks, in higher dimensions it is necessary to resort to either some approximate contraction scheme, or to statistical, Monte-Carlo type methods. 

A possible strategy is to place additional restrictions upon the tensor network that render the quantities of interest easier to calculate. Such restrictions involve a compromise in accuracy, which must be balanced against gains in efficiency. Here, we consider the set of states at the intersection between MPS and CPS. The subset of CPS in which each spin is contained in only two plaquettes are also a subset of PEPS - the latter restricted to states with the same bond dimension and local Hilbert space dimension. 
Examples include stabiliser states \cite{Gottesman:1997zz} and Kitaev's toric code \cite{Kitaev:1997wr}. In common with string bond states \cite{2008PhRvL.100d0501S}, of which they are a subset, CPS are efficiently sampleable, but not in general efficiently contractible, and can be used to calculate expectation values of local observables efficiently.  The trade-off is that CPS are basis dependent - meaning that CPS defined with respect to different bases do not cover the same submanifold of the Hilbert space.  Physical insights can motivate a judicious choice of basis for a given Hamiltonian.  

We focus upon the above-mentioned subset of uniform CPS (uCPS), in one spatial dimension, and in the thermodynamic limit. We have two main aims:

\noindent
i. {\it To develop the deterministic methods for optimising and evolving uCPS}. Our main tool will be a deterministic algorithm based upon the time-dependent variational principle (TDVP), an approach that projects the exact Schr\"{o}dinger time-evolution onto the uCPS sub-manifold of the full Hilbert space.  The deterministic TDVP approach is somewhat orthogonal to the usual statistical setting in which CPS
are utilised.\footnote{Although other deterministic methods have been developed in \cite{Changlani:2009mz} and \cite{Neuscamman:2011lq}.} This brings with it certain advantages, such as a very robust algorithm, as well as the ability obtain time-evolution information straightforwardly; but the main disadvantage is that with current techniques we are restricted to the one-dimensional setting. Nevertheless, many of the results provide insight about higher dimensional behaviour. Our construction will follow closely the TDVP implementation for MPS given in \cite{2013PhRvB..88g5133H}.

\noindent
ii. {\it To provide a systematic study of basis dependence of uCPS and its interplay with computational speedup.} Speedup is expected due to the due to the ability to  store and manipulate uCPS matrices more efficiently than their unrestricted uMPS counterparts, but this advantage is offset by the lower capacity of uCPS to encode ground state and quench information. Indeed, the interplay that we reveal is more subtle than one might anticipate.

\section{Outline}
The basics of (u)CPS are reviewed in Section \ref{sec:CPS}, and their relation to (uniform) matrix product states (u)MPS is derived. The (u)CPS $\rightarrow$ (u)MPS mapping necessitates the introduction of the so called copier tensor. This construction is not sensitive to our one dimensional analysis, and it is the structure of the copier tensor in higher dimensions that is the starting point, and can provide indications, of how our one-dimensional results can be generalised.

In Section  \ref{sec:uCPS_TDVP}  the time-dependent variational principle (TDVP) for uCPS is derived.  Since the cost of a single uMPS TDVP step scales as $O(d D^3)$, where $d$ is the size of the physical and $D$ of the virtual dimension, one naively expect the cost of each TDVP step to scale as $O(D^4)$, but we demonstrate that for uCPS the cost drops to $O(D^3)$.    

Section  \ref{sec:properties_of_uCPS_gs} provides a detailed study of  the properties of uCPS ground state approximations, exemplified by a number of models: the Quantum Ising-, the Heisenberg-, and the XY-model (all spin-$\frac{1}{2}$).  The imaginary time algorithm converges to the global minimum uniquely only for Hamiltonians that do not have degenerate groundstates. When the groundstate is degenerate it may still be possible to find  a special choice of basis for which convergence is unique, however in general it turns out that uCPS breaks the degeneracy due to its basis dependence, and TDVP may converge to local minima associated with this breaking.  When convergence is not unique, the algorithm acquires a probabilistic ingredient. For a Hamiltonian with a discretely degenerate groundstate the computational cost is increased only by a constant factor, but if the vacuum has a continuous degeneracy this factor seems to be larger than constant, and the number of possible local minima that TDVP can converges to seems to increase in an unbounded manner with bond dimension.  
Next, the efficiency of uCPS is compared with uMPS. We uncover an intricate picture.  Naively one might expect that the uCPS in its optimal basis would capture a state with accuracy comparable to uMPS of which it is a restriction - {\it i.e.} uMPS with bond order equal to the local Hilbert space dimension. In fact, the situation is considerably worse; the plaquette overlap required to match the accuracy of a uMPS of a given bond dimension is proportional to that bond dimension. This is exponentially worse than naive expectation since the local Hilbert space dimension scales exponentially with the size of the overlap region. However, in the optimal basis, we find that  the computer \emph{time} needed to converge the ground state approximation to a desired accuracy scales in the same way for both uCPS and uMPS. We also study the scaling of quantities, such as energy, entanglement entropy, or the correlation length, with bond dimension in detail. The scaling behaviour  is generally basis dependent, but much smoother for uCPS than for uMPS. This allows for a more accurate estimates of certain physical quantities to be made with uCPS than with uMPS with the same computational  time cost (but, still, at exponentially larger computer memory cost). In general we find that an optimal basis corresponds to one that is aligned  as closely as possible with the entanglement generating terms in the Hamiltonian.\footnote{This is clearly only a loose guiding principle, discussed at length in Section  \ref{sec:properties_of_uCPS_gs}, and exemplifies the issue one needs to address for CPS/string bond states in general, of identifying the optimal basis for the physics one wishes to study.} 

In Section \ref{sec:CPS_quenches}, real time TDVP is applied to uCPS in order to simulate quantum quenches. We concentrate in particular upon quenches across a critical point that exhibit dynamical phase transitions. It is found that the capacity of uCPS to correctly capture a quench to a desired accuracy scales, analogously to the static context above, exponentially worse than uMPS in memory requirements but in the same manner as far as time costs are concerned. This is again only true provided one is working in the optimal basis, or reasonably close to it; but unlike the static case, here there are no probabilistic aspects to contend with.   It is also found that for a sub-optimal basis choice, the uCPS approximation to the quench can completely miss the dynamical phase transition. In fact, with the same basis choice, the uCPS approximation of the ground state misses the equilibrium phase transition. All of this gives some indication of the types of issues one may encounter with CPS in general, if working in a sub-optimal basis.
 
We elaborate upon these issues in Section \ref{sec:discussion} and give a final brief summary in Section \ref{sec:conclusions}. To keep things as self-contained as possible, a basic review of MPS and TDVP is provided in Appendix \ref{app:MPS_review}. Appendix \ref{app:uCPS_contraction} details the contraction ordering that achieves the optimal  $O(D^3)$ scaling of a single uCPS TDVP step, while details of a preconditioning step necessary for an iterative sub-routine in the TDVP algorithm to scale optimally is provided in Appendix \ref{app:uCPS_precond}.

\section{Correlator Product States}
\label{sec:CPS}

In this section the basics of correlator product states (CPS) and  the relationship of the CPS variational class to string bond states will be reviewed. We then focus upon the main subject of this paper, uniform correlator product states (uCPS) in the thermodynamic limit, detail how the uCPS class can be understood as a subset of uniform matrix product states (uMPS), and discuss a number of implications. For a brief review of Matrix Product States (MPS) the reader is referred to Appendix \ref{app:MPS_review}.

\noindent \emph{Definition of CPS - } A CPS wavefunction on an $N$-site spin lattice is formed by dividing the lattice into a set of overlapping subsets called plaquettes - labelled here by $P$. Wavefunctions are defined over each plaquette by a tensor $C_P$ and a correlated product between them taken according to 
\begin{align}
\label{eq:CPS_def}
\ket{ \Psi [ C ] }= 
\sum_{i_1, i_2 \cdots i_N } \prod_P C_P^{i_1, \cdots i_{l} } \ket{ i_1, i_2, \cdots, i_N }  \  \ 
\end{align}
in order to maintain consistency in the regions of overlap and thereby entangling spins in disjoint plaquettes. The plaquette sizes are taken to be independent of the size of the system, and sufficiently small that the wavefunction on each can be stored and manipulated with small computational cost. The accuracy of this Ansatz is influenced by a large number of factors,  including the structure of the Hamiltonian, the choice of basis, and size and number of overlaps between the plaquettes.

\begin{figure}[!htb]
\centering
\includegraphics[width=9cm]{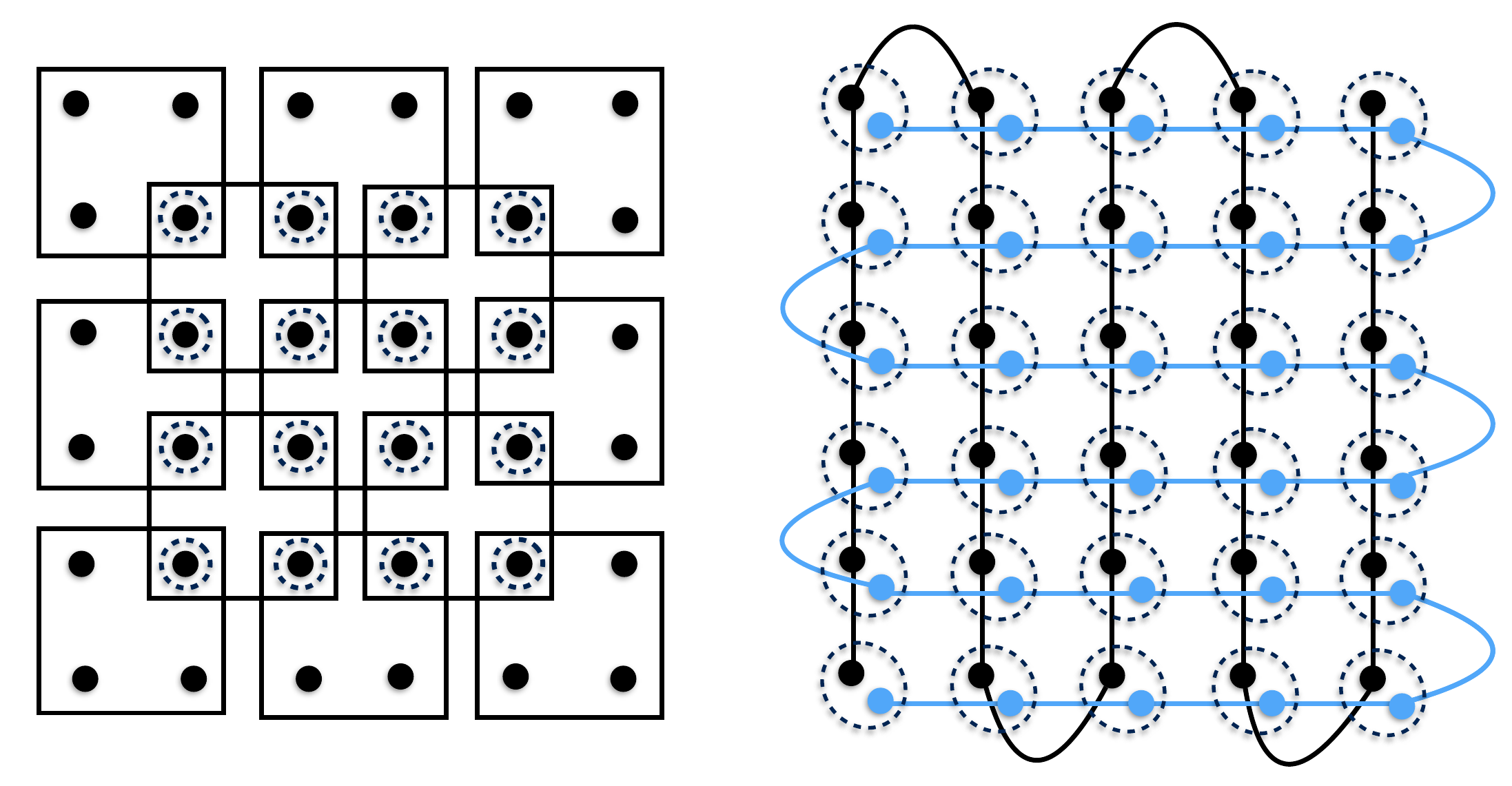}
\caption{\footnotesize{ (Colour online).  The diagram on the left hand side depicts a CPS in two dimensions with single-site overlaps. The squares represent the $C$ tensors, and the black dots denote the spins, each $C$ tensor thus having four indices. The right hand side depicts a string bond state consisting of a product of two matrix product states, each covering the entire lattice. The  dashed circles are present at sites where the underlying CPS/string bond state tensors overlap, and each such circle can be understood as representing the copier tensor projecting the indices in the overlap onto the physical spin. } }\label{fig:CPS_string_bond_2D_diagram}
\end{figure} 

The key requirement for the Ansatz of the form (\ref{eq:CPS_def}) to be amenable to Monte Carlo type algorithms is that the coefficients of the wavefunction $\ket{\Psi [ C ]}$ in (\ref{eq:CPS_def}) be efficiently calculable.\footnote{This is only a necessary condition, and is not sufficient to guarantee that the Monte Carlo algorithm will converge.}  This is clearly satisfied whenever the coefficients  $C_P$ are efficiently calculable individually.  For CPS this requirement holds simply because the plaquettes are by construction taken to be of a small constant size, but more generally \emph{any} efficient format for $C_P$ will be a priori equally suitable.  The definition of the class of \emph{string bond states} \cite{2008PhRvL.100d0501S} is also of the general form (\ref{eq:CPS_def}), but with the sub-lattices $P$ taken to be strings of arbitrary length instead of plaquettes, and the $C_P$ coefficients given by a matrix-product form.  Examples of both  a CPS with single-site overlap, and a string bond state given by a product of two MPSs, each covering the entire lattice, are depicted in Figure \ref{fig:CPS_string_bond_2D_diagram}.

Since a general-form wavefunction on a constant size plaquette can always be recast in MPS form, CPS clearly form a subset of string bond states. It is, however, not the case that in general either CPS or String Bonds States can be efficiently represented as PEPS (or MPS in one dimension). The number of overlapping plaquettes for a CPS, or strings for a string bond state, is allowed to be of the order of system size, yet representing such a state as PEPS/MPS requires in general  an exponentially large bond dimension.   The Laughlin state, for example, can be written as a 2-site CPS \cite{2009PhRvB..80x5116C}, but the number of overlapping CPS tensors at each site is equal to the system size; a diagram depicting the general form of the Laughlin state is given in Figure \ref{fig:CPS_Laughlin_4_site}. 

A one dimensional CPS with open boundary conditions, and with the restriction that only two plaquettes overlap, is given by:
\begin{align}
\label{eq:CPS_def_1D}
& \ket{ \Psi [ C ] }= \\ \nonumber & \sum_{i_1, i_2 \cdots i_N } v_L^{i_1} C_{1}^{i_1 i_2 } C_{2}^{i_2 i_3 } \cdots C_{N-1}^{i_{N-1} i_N } v_R^{i_N} \ket{ i_1, i_2, \cdots, i_N }  \ .
\end{align}

\noindent \emph{ The copier tensor - }    In order to write down a tensor network for a CPS it is necessary to introduce the copier tensor. For the state (\ref{eq:CPS_def_1D}), for example, the copier is given by $\eye_{ijk}$,  defined as the tensor with components  equal to unity when all its indices are equal, and zero otherwise.  It will be convenient to represent the copier graphically as \includegraphics[scale=0.1]{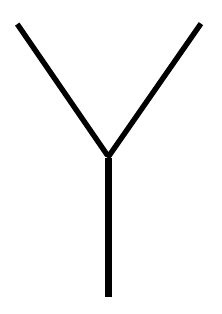};  the CPS state in (\ref{eq:CPS_def_1D}) can then be represented as:   \newline
\centerline{ \includegraphics[scale=0.3]{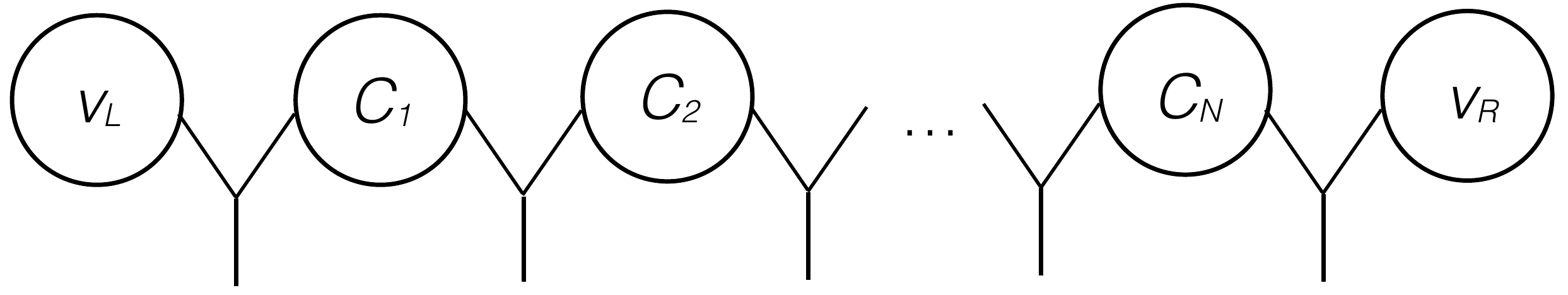}  \ \ \raisebox{+2ex}{,} } 
 \newline
where any variation of the dimensionality of the indices in (\ref{eq:CPS_def_1D}) along the chain can be indicated by an additional label if necessary.  As each physical index in (\ref{eq:CPS_def_1D}) corresponds to the overlap of the corresponding CPS plaquette, for an $n$-site overlap the dimensionality of the index is obtained by  grouping the $n$ fundamental spins $s$ together, and is given by $s^n$.  The dimensionality of these indices therefore grows exponentially with the size of the overlap. The basis dependence of CPS/string bond states is made explicit by the non-covariant definition of the copier tensor, and it is easily verified that a rotation by a local unitary on the physical leg can not in general be "pulled through" to the virtual level.\footnote{One can consider adding such local unitaries, or more general matrices, in order to enlarge the CPS variational class and restore rotation independence. }

The copier tensor generalises straightforwardly to more general CPS, and to  higher dimensions.  For example, the dashed circles in Figure \ref{fig:CPS_string_bond_2D_diagram} should in fact be understood as representing three-index copier tensors. For the Laughlin state depicted in Figure \ref{fig:CPS_Laughlin_4_site}, the dashed circles again represent copiers, but the number of virtual legs is now of the order of system size. 

\begin{figure}[!htb]
\centering
\includegraphics[width=6cm]{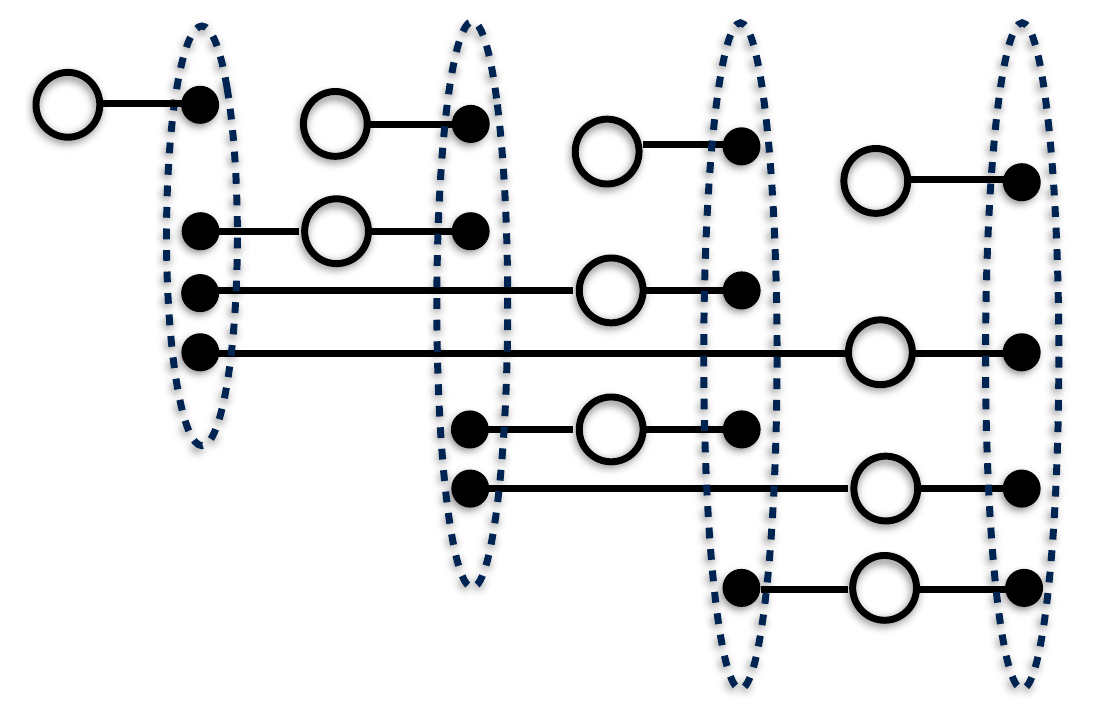}
\caption{\footnotesize{ This diagram depicts the general form of a tensor network corresponding to the CPS representation of the Laughlin state over four spins, and exemplifies a CPS for which the number of overlapping CPS tensors at each physical spin is of the order of system size. The circles denote the CPS tensors, with their indices depicted by lines ending in the black dots, and the dashed ovals represent the action of the copier tensors mapping overlaps of the CPS tensors onto the four physical spins. The diagram corresponding to the CPS representation of a Laughlin state over an arbitrary number of spins follows the same pattern.    } }\label{fig:CPS_Laughlin_4_site}
\end{figure} 

Furthermore, using the tensor network representation of CPS, a natural generalisation of the CPS/string bond state variational class can be achieved by promoting the copier  to any tensor that is sufficiently sparse to leave sample-ability intact.  As an example, one can imagine a CPS-type Ansatz for which the copier is replaced by a tensor the non-zero components of which are given by an $n$-site MPS.

\noindent \emph{ Mapping one-dimensional CPS into MPS - }   One dimensional CPS of the form (\ref{eq:CPS_def_1D})  form a sub-class of MPS with open boundary conditions,
\begin{align}
\label{eq:MPS_def}
& \ket{ \Psi [ A ] }= \\ \nonumber & \sum_{i_1, i_2, \cdots, i_N}^d  v_L^\dagger A^{i_1}_1 A^{i_2}_2 \cdots A^{i_N}_N  v_R \ket{ i_1, i_2, \cdots, i_N }  \  ,
\end{align}
provided that the physical spin- and bond-dimensions  of the MPS tensors, which at site $K$ we denote as $d^{(K)}$ and $D^{(K)}$ respectively, are the same either to the left or to the right (i.e. provided that either $d^{(K)} = D^{(K-1)}$, or $d^{(K)} = D^{(K)}$).  There are many ways to map the CPS defined in (\ref{eq:CPS_def_1D})  into an  MPS of this type.  One possibility is the following: \newline
\centerline{  \includegraphics[scale=0.3]{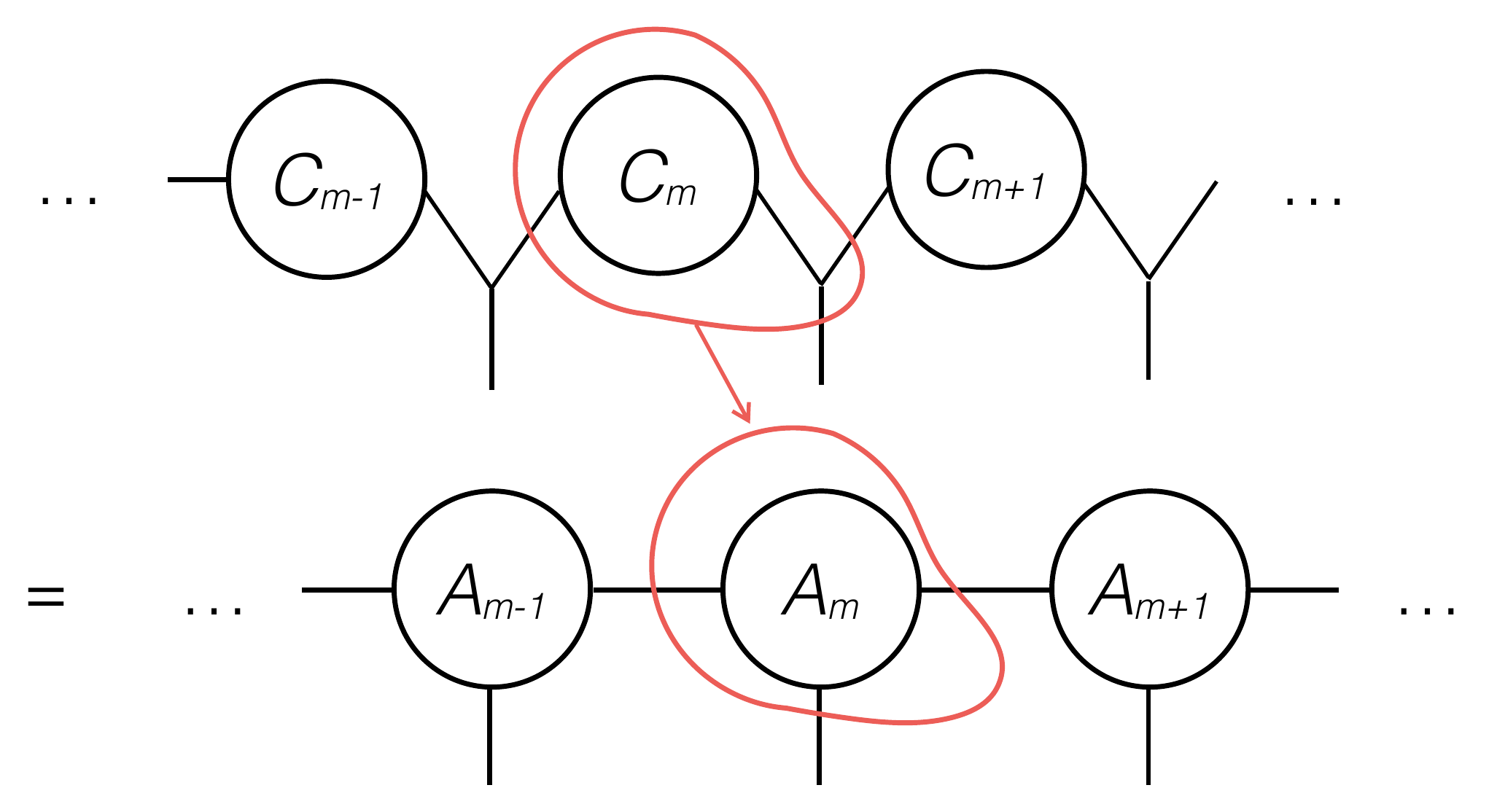} \ \ \raisebox{+2.5ex}{.} }  \newline
  Alternatively the $C$ matrix can be placed to the right of the copier in the identification, or placed both to the left and the right of the copier after decomposing each $C$ matrix into a product of two matrices. 
  
In higher dimension, any CPS/string bond state with a constant number of overlapping plaquettes/strings can be mapped to a PEPS with small bond dimension using analogous considerations.
  
\noindent \emph{  Uniform Correlator Product States (uCPS) - }   This paper is concerned mainly with uniform translation invariant CPS in the thermodynamic limit. In order to achieve a fully translation invariant representation, it is necessary that the plaquettes contain an even number of spins, so that half of the spins of each plaquette overlap with half of the spins of a neighbouring plaquette.\footnote{More generally one can consider representation that are invariant under translations by $m$ sites, with $m>1$.}   This means that the $C$ matrices are square and that the bond dimension along the chain is constant. These states will be referred to as uniform correlator product states (uCPS), and form a proper subset of the uMPS class described in Appendix \ref{app:MPS_review}.  Graphically a uCPS is represented as: \newline
\centerline{ \includegraphics[scale=0.3]{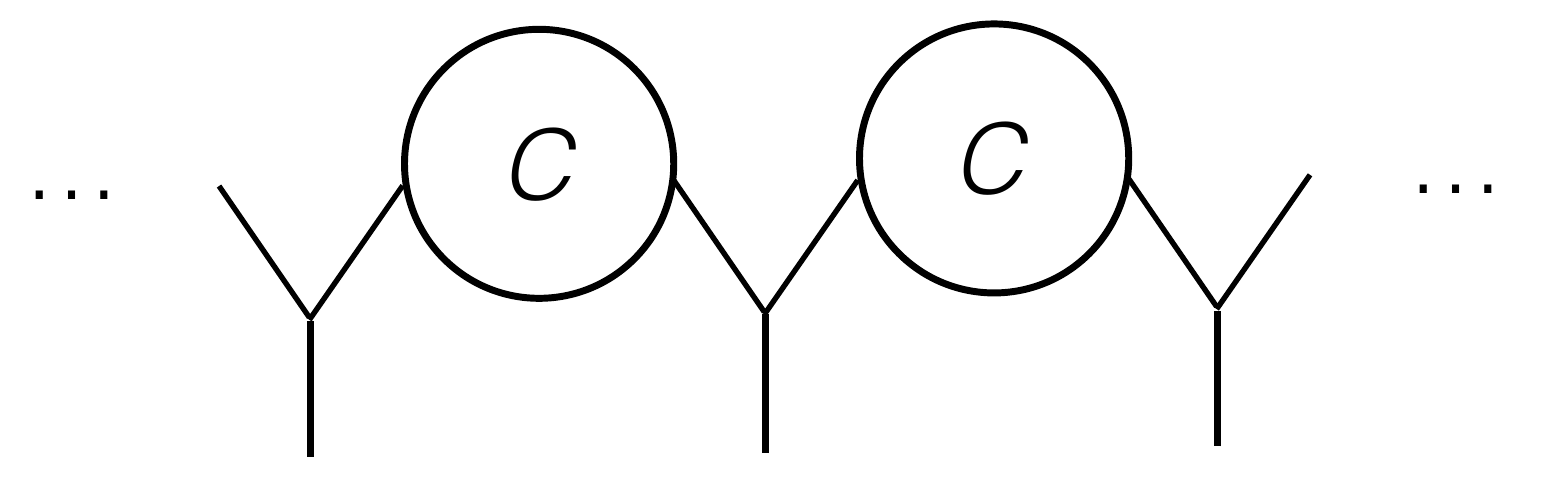}  \ \ \raisebox{+3ex}{.} } \newline
The uCPS $\rightarrow$ uMPS mapping  is achieved in the same manner as in the non translation invariant setting described above, with analogous freedom in how this mapping is realised.   

The increase in uCPS bond dimension is achieved by increasing the size of the overlap between plaquettes.  If $n$ is the number of fundamental spins in each overlap, for a lattice of  $s$-level spins the size of the C matrices is $s^n \times s^n$, and the bond dimension $D$, given by $D = s^n$, therefore increases exponentially with $n$. 

\noindent \emph{The uCPS transfer matrix - } The transfer matrix $\tilde{E}$  (see equation (\ref{uMPS_transfer_matrix}))  corresponding to the uCPS $\rightarrow$ uMPS mapping with the $C$ matrix placed to the right of the copier tensor takes the form:
\newline
{ \centerline{  \includegraphics[scale=0.3]{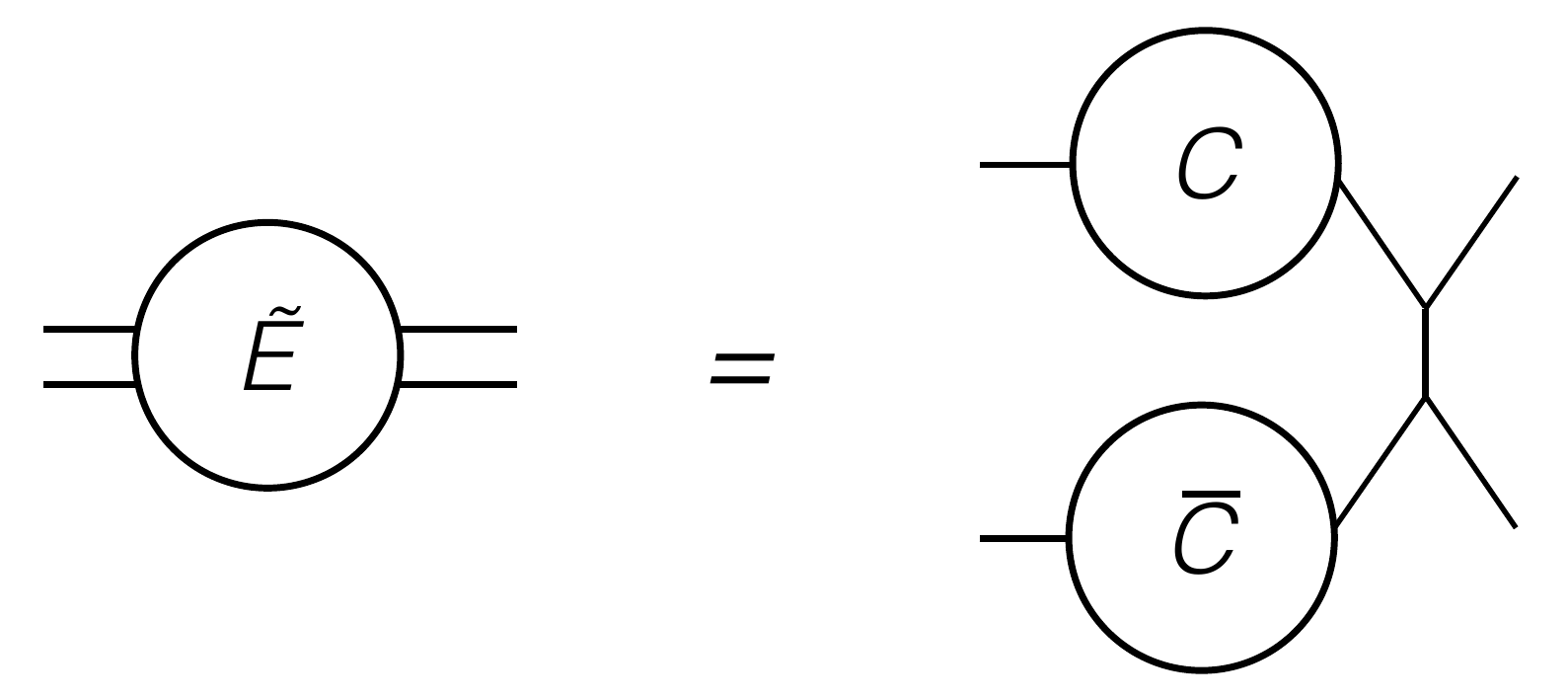}  \ \ \raisebox{+4ex}{.}  }  }\newline
Due to the sparseness of the copier tensor, $\tilde{E}$ has only $D$ non-zero eigenvalues. This is related to the fact that the the object  \includegraphics[scale=0.08]{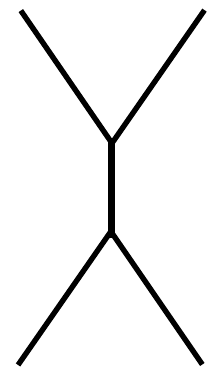} is a projector that effectively reduces the size of a $D^2$ dimensional index to a $D$ dimensional one. It also follows from this that the  uCPS variational class is in fact a subclass of reduced rank - not full rank -  uMPS. It is sufficient therefore to work with the projected version of the above object, $E$: \newline
{ \centerline{  \includegraphics[scale=0.3]{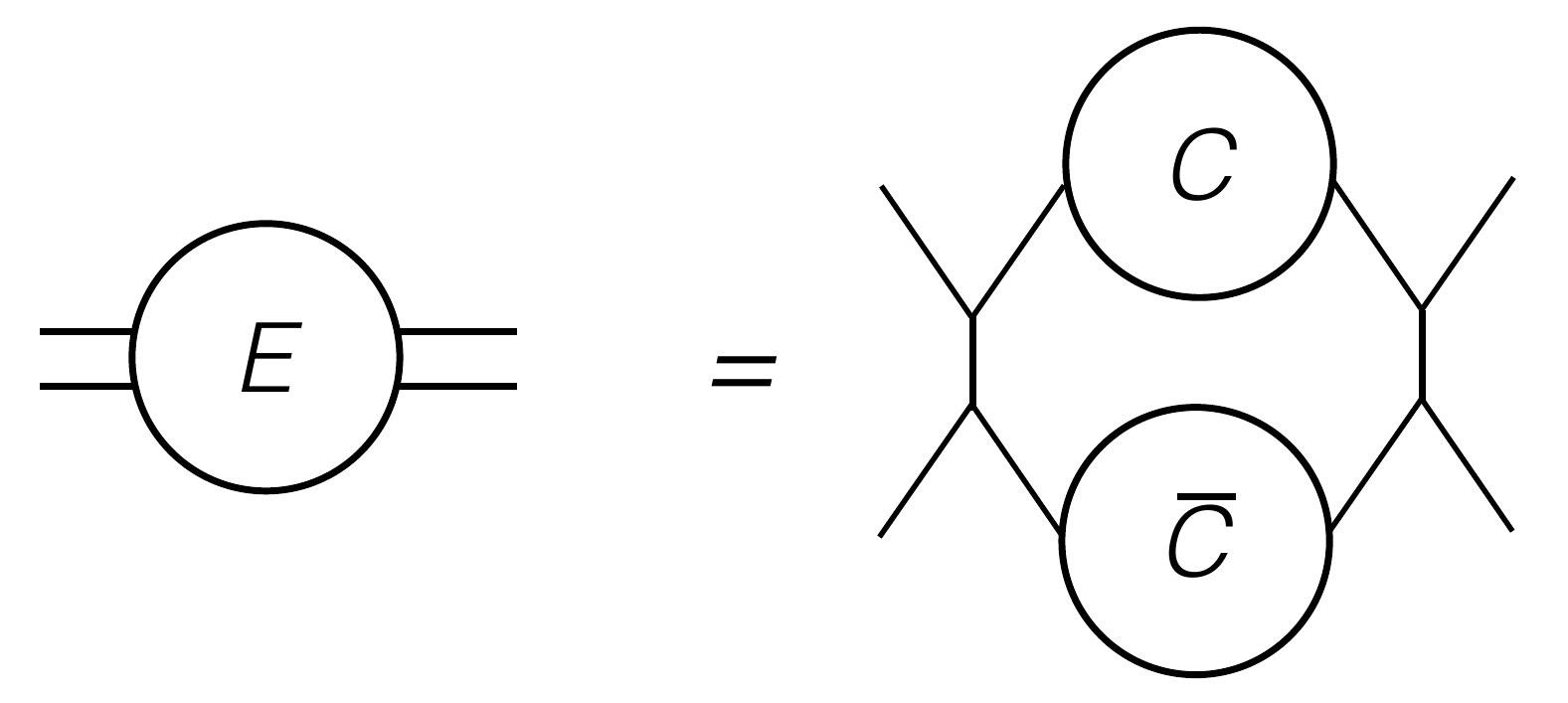} \ \ \raisebox{+4ex}{,}   }   }\newline
the non-zero subspace of which is given by the $D \times D$ dimensional matrix:
\begin{align}
\label{eq:uCPS_transfer_matrix}
E_{ ij} =  C_{ij} (.*) \overline{C}_{ij}  \ .
\end{align}
The operator $(.*)$ denotes component-wise multiplication for repeated indices. As for uMPS, in order to ensure finite normalisation of the state, it is necessary to normalise the $C$ matrices so that the largest eigenvalue of $E$ is one. 

As discussed in Appendix \ref{app:MPS_review},  the left and right eigenvectors of the transfer matrix $\tilde{E}$ corresponding to eigenvalue one are centrally important objects,  which determine the Schmidt coefficients across a cut of the infinite chain into two semi-infinite intervals, and, more generally, represent the cumulative effect of the environment when calculating the expectation value of a local operator, from its insertion to $\pm \infty$. In order to fully utilise the special form of uCPS over a generic uMPS in the TDVP algorithm (see Section \ref{sec:uCPS_TDVP}),  it is necessary to work directly with the  $D$ dimensional left and right eigenvectors of $E$  (\ref{eq:uCPS_transfer_matrix}) corresponding to eigenvalue one (rather than with the eigenvectors of $\tilde{E}$), which we will refer to as $V_L$ and $V_R$, respectively (these should not be confused with the boundary $v_L$ and $v_R$ of open-boundary finite MPS (\ref{eq:MPS_def})).   We note that, while the left eigenvector of $\tilde{E}$ is given by:
\begin{align}
\label{eq:uCPS_left_eigv}
{ \centerline{  \includegraphics[scale=0.3]{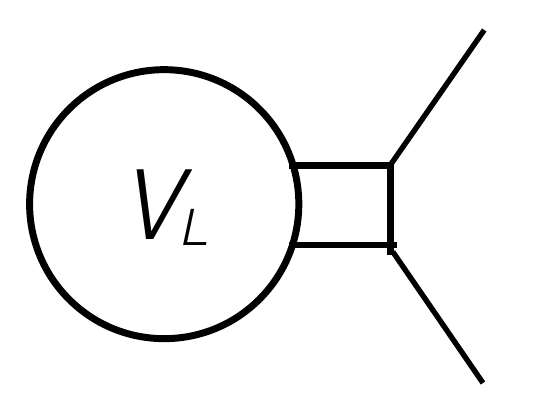}  \ \ \raisebox{+2.5ex}{,}  } } 
\end{align}
  the right eigenvector explicitly depends on $C$  as:
  \begin{align}
 \label{eq:uCPS_right_eigv} 
 { \centerline{  \includegraphics[scale=0.3]{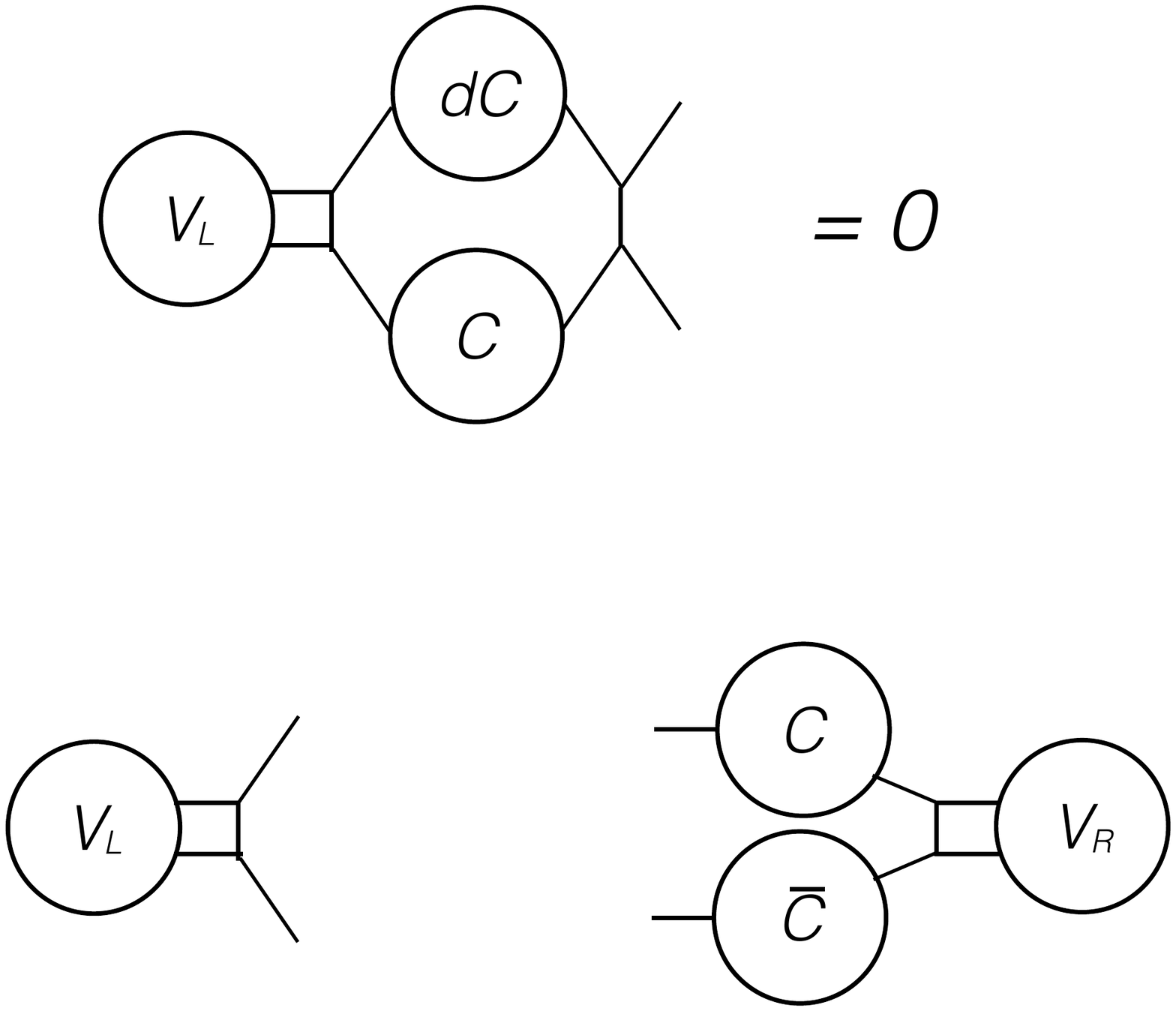} \ \ \raisebox{+3ex}{.}   } } 
 \end{align}
Had we made the identification with uMPS by placing the $C$ matrices to the right of the copier tensor, the situation would be the reverse of the above, with the right eigenvector depending  only on $V_R$, and the left on $V_L$, $C$, and $\overline{C}$.

A uMPS can always be gauge-transformed (\ref{eq:uMPS_gauge_transf}) to the left or right canonical gauge, as discussed in Appendix \ref{app:MPS_review}, meaning that either the left or right environment matrix is equal to the unit matrix.  This is in general not achievable with uCPS, since the set of uMPS gauge transformations that preserve the uCPS form is extremely restricted, and the remaining gauge freedom is not sufficient in general to achieve the left or right canonical gauge. Stated in another manner, uCPS corresponds to a very restricted type of uMPS, and  imposing such a restriction on uMPS fixes nearly all of its gauge freedom. One of the implications of all of this is that it does not seem possible, at least straightforwardly, to implement an analogue of the  iDMRG algorithm \cite{2008arXiv0804.2509M} for uCPS, since it relies upon the ability to achieve the left and right canonical gauges and to switch between them.  It is interesting that there are nevertheless two natural gauge choices for uCPS, as explained above, related to the freedom of placing the copier to the left or to the right of the $C$ matrix in the uCPS $\rightarrow$ uMPS mapping. 
 
 \noindent \emph{Computational cost of contracting uCPS - } Since a uCPS maps to a uMPS with equal bond and physical dimensions, i.e. $d=D$, and the cost of calculating a local uMPS observable is $O(d D^3)$ (see Appendix \ref{app:MPS_review}), one naively expects that the cost of calculating expectation values of local observables for uCPS scales as $O(D^4)$. It turns out, however, that one can do better.  One possible reduction in cost occurs due to the fact that the copier tensor factorises, and that a uCPS with overlap of size $n$ and local spin of dimensionality $s$, can be mapped to an $n$-site\footnote{A $n$-site uMPS is a MPS in the thermodynamic limit such that its tensors are unchanged after a translation by $n$ sites.} uMPS with bond dimension $D= s^n$.  How this is achieved is most clearly demonstrated graphically. For example, for $n=3$ and $D=s^3$ the uMPS tensor of the uCPS form can be decomposed into $3$ MPS tensors as:
 \newline
{ \centerline{  \includegraphics[scale=0.3]{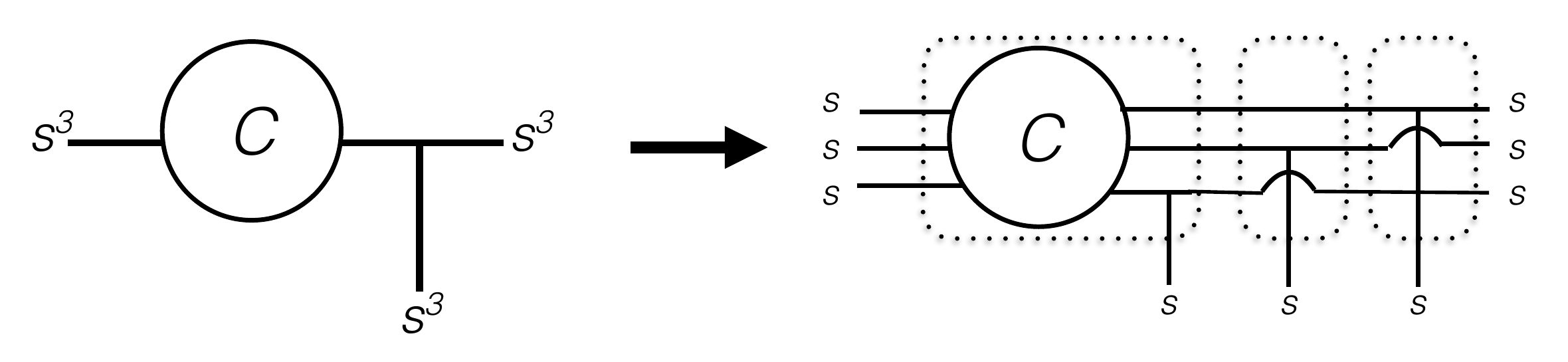}  \ \ \raisebox{+1ex}{,}  } }  \newline 
  where the dashed regions identify the three MPS tensors with physical dimension $s$.\footnote{Clearly all the uCPS information has been incorporated into the left-most tensor for this particular choice of decomposition, but if desired $C$ can be distributed over the $n$ MPS tensors in a symmetrical manner by performing a suitable factorisation.}  The cost of calculating local expectation values for an $n$-site uMPS can therefore be reduced to $O( \log(D)  s D^3) = O( \log(D)  D^3)$, simply by applying an optimal contraction ordering for $n$-site uMPS after making the above decomposition. 
  
The presence of the $D^2 \rightarrow D$ projector \includegraphics[scale=0.08]{uCPS_projector}  in the transfer matrix $\tilde{E}$ hints that an even lower cost may be achievable. This turns out to be the case, and in the next section we demonstrate that, in addition to the computation of expectation values of local operators, all the steps of the uCPS TDVP algorithm can be computed with cost $O(D^3)$.

\section{TDVP with \MakeLowercase{u}CPS}
\label{sec:uCPS_TDVP}
  The time-dependent Schr\"odinger equation, $\frac{d}{dt} \ket{ \Psi (t) }  = -i \hatH (t) \ket{\Psi (t)}$,
does not in general have an exact solution if the state vector is restricted to remain within some class of variational states $\ket{ \Psi (V_X) }$ during the evolution, where $V_X$ denotes some set of variational parameters labelled by an index $X$. This can be observed by considering the equation:
\begin{align}
\label{eq:var_vs_exact}
\sum_X \dot{V}_X \frac{\partial}{\partial V_X } \ket{ \Psi (V_X) } + i \hatH (t) \ket{\Psi (t)}  = 0\ ,
\end{align}
where the second term can correspond to an arbitrary direction in Hilbert space, but the first is highly restricted.  Instead, one must adopt some procedure for obtaining an optimal $\dot{V}_X$ that minimises the left hand side of (\ref{eq:var_vs_exact}) with respect to some cost function.   The time-dependent variational principle (TDVP) corresponds to the  natural cost function in a quantum mechanical setting, namely the $2$-norm, and the  solution to $\dot{V}^*_X$  is given by: \footnote{See \cite{PhysRevLett.107.070601, 2013PhRvB..88g5133H} and Appendix \ref{app:MPS_review} in reference to the application of TDVP to (u)MPS.}
 \begin{align}
 \label{eq:TDVP_min_general}
  \dot{V}^{*}_X =  \argmin_{\dot{V}_X} \norm{ \sum_X \dot{V}_X \frac{\partial}{\partial V_X } \ket{ \Psi (V_X) }  +  i \hatH (t) \ket{\Psi (t) } }^2 \ .
 \end{align}

 The full variational space of uCPS  consists of a general variation $dC$ of the uCPS matrix $C$. In the context of the uCPS $\rightarrow$ uMPS mapping described in Section \ref{sec:CPS}  this corresponds to a restricted uMPS variation of the form:
 \begin{align}
 \label{eq:dC_uMPS_variation}
\centerline{ \includegraphics[scale=0.3]{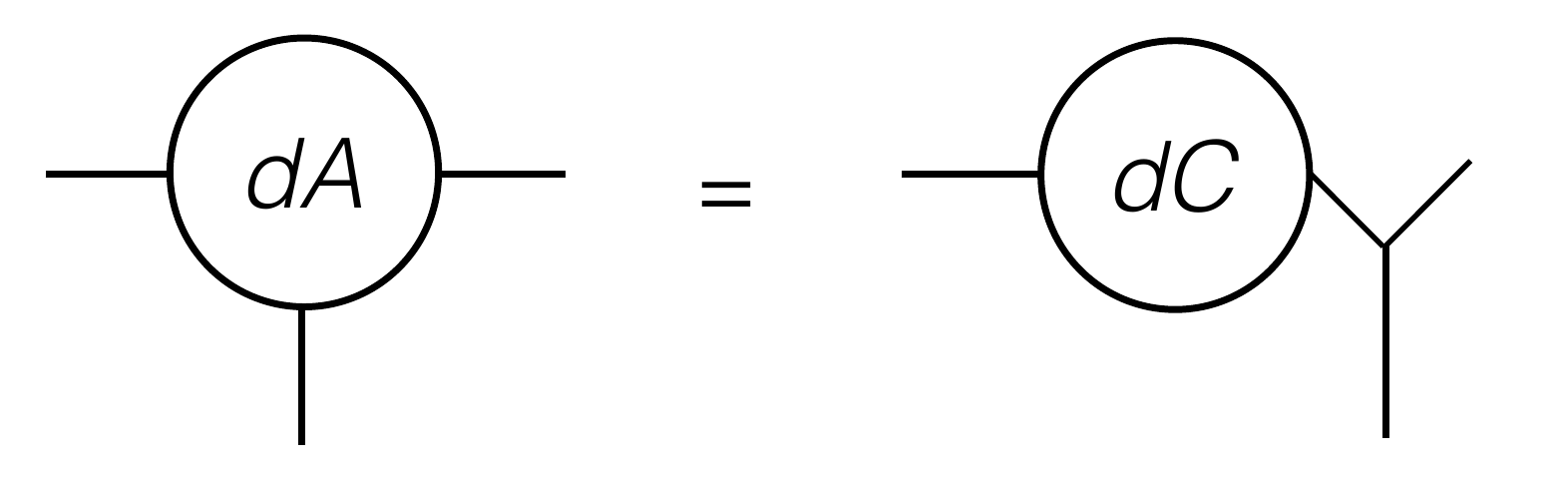}  \ \  \raisebox{+2.5ex}{,}  }
\end{align} 
and can be understood as a tangent vector to the uCPS manifold (see also Appendix \ref{app:MPS_review}). Writing this out symbolically: 
\begin{align}
\label{eq:uCPS_tangent_vec_general}
 \ket{ \Phi(dC, C)} := \sum_{i, j} dC_{ij} \frac{\partial}{\partial C_{ij}} \ket{\Psi(C)}  \equiv  dC^{ij }  \ket{ \partial_{ij} \Psi(C)} \ ,
 \end{align}
the solution to the minimisation problem (\ref{eq:TDVP_min_general}) is given by:
 \begin{align}
\label{eq:uCPS_TDVP_equation_general_dC}
\sum_{k, l} G_{\overline{ij} |   kl}  \dot{C}^{kl} = & -i \bra{ \partial_{\overline{ij}} \Psi(  C(t) )  } \hatH (t) \ket{\Psi  (C(t) ) }   \ ,
\end{align}
where $G$ is the uCPS Gram matrix (c.f. (\ref{eq:uMPS_gram_matrix})):
\begin{align}
\label{eq:uCPS_gram_matrix}
G_{\overline{ij } |  kl} := \braket{\partial_{\overline{ij}} \Psi(  C) }{\partial_{kl}  \Psi(C)     }  \ . 
\end{align} 
  
However,  the solution in (\ref{eq:uCPS_TDVP_equation_general_dC}) contains terms that diverge if the variations $dC$ are left fully unconstrained.  The full tangent space spanned by $\ket{ \Phi(dC, C)}$ includes transformations along the state $\ket{\Psi(C)}$ itself, and it is precisely such norm-changing variations that generate divergences.  These must be projected out, which can be achieved by implementing the uCPS analogue of the uMPS left or right tangent space gauge condition (\ref{eq:uMPS_l_r_tangent_gauge}).  Moreover, the tangent space gauge condition simplifies the form of the Gram matrix, which becomes local, meaning that all the contributions to  (\ref{eq:uCPS_gram_matrix}) vanish except those for which the variations $dC$ in the bra and ket occur at the same lattice site. 

\noindent \emph{Gauge fixing for uCPS - }
For the sake of concreteness we will demonstrate how to impose  the \emph{left} tangent gauge condition for uCPS, which reads: \newline
\centerline{ \includegraphics[scale=0.3]{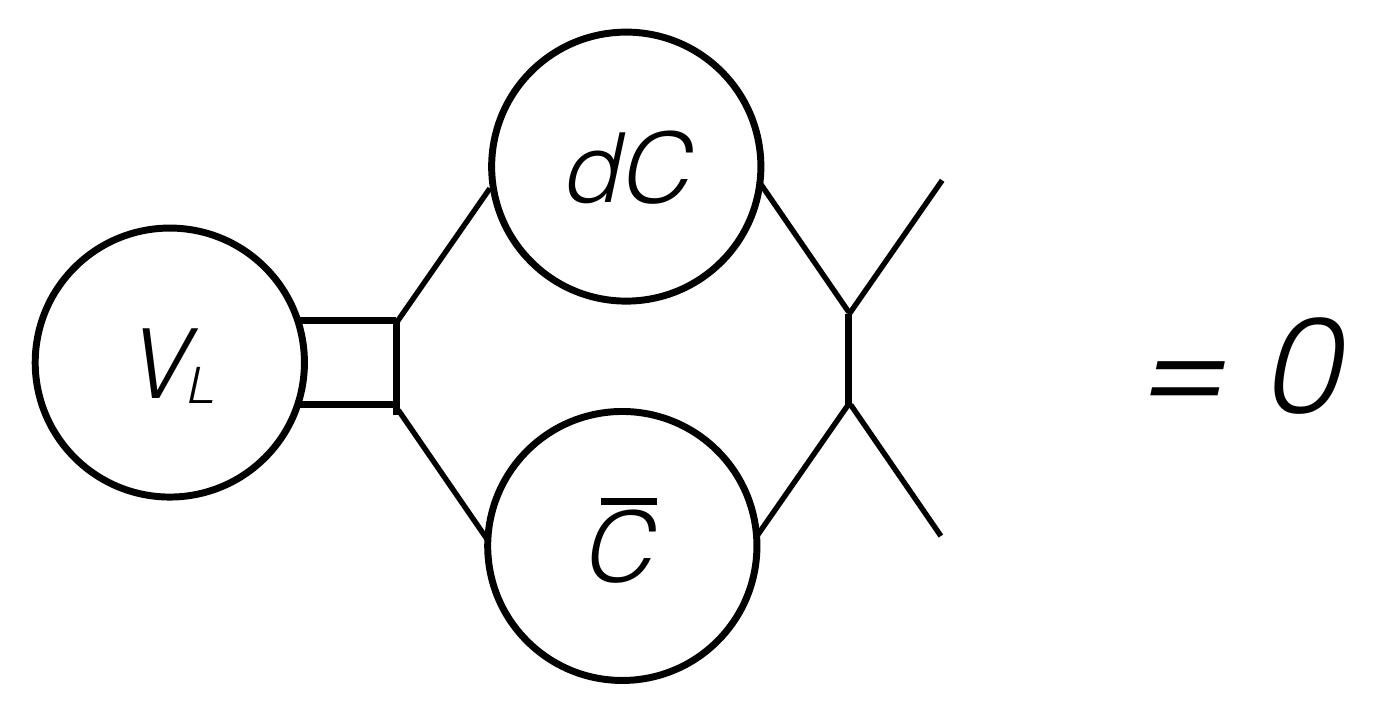}  \ \ \raisebox{+3ex}{,}  } 
\newline
or symbolically
\begin{align}
V_{(L )i} (.*) dC_{i j} (.*) \conj{C}_{ij} = 0 \ .
\end{align}
A solution to this equation is given by $dC$ of the form:
\begin{align}
\label{eq:dC_uCPS_TDVP}
dC_{ij} = \sum_{\tilde{\alpha} \beta} B_{\tilde{\alpha} \beta} V_{(R)  i }^{\tilde{\alpha}} V_{(L)  j}^{\beta} (.*) ( 1 (. /) \conj{C}_{  i j}) \ \ ,
\end{align}
where $B$ is an arbitrary $(D-1) \times D$ matrix, $(./)$ denotes component-wise division, and $V_{(L)}^\alpha$ and $V_{(R)}^\alpha$ are the $D$ left and right eigenvectors of the transfer matrix (\ref{eq:uCPS_transfer_matrix}), so that
\begin{align}
E=  \sum_\alpha \lambda_\alpha V_{(L)}^{\alpha} V_{(R)}^{\alpha} \ .
\end{align}
By convention $V_{(L)}^{1} \equiv V_{(L)}$,  and $V_{(R)}^{1} \equiv V_{(R)}$,  and the  $\tilde{\alpha}$ index in (\ref{eq:dC_uCPS_TDVP}) thus runs from $2$ to $D$. 

With the tangent space gauge condition in place, the analogue of (\ref{eq:uCPS_TDVP_equation_general_dC}) for $B$ is given by:
\begin{align}
\label{eq:uCPS_TDVP_equation}
\sum_{\tilde{\gamma} \delta } G_{\overline{\tilde{\alpha} \beta} |  \tilde{\gamma} \delta }  B^{\tilde{\gamma} \delta} = & -i \bra{ \partial_{\overline{ \tilde{\alpha} \beta }} \Psi(  C(t) )  } \hatH (t) \ket{\Psi  (C(t) ) }  \ ,
\end{align}
where $\partial_{\overline{ \tilde{\alpha} \beta }}$ on the right hand side stands for a derivative with respect to $\overline{B}^{\tilde{\alpha} \beta}$, after making use of the chain rule applied to a differential  $dC$ of the form (\ref{eq:dC_uCPS_TDVP}).  After deriving the solution for $B_{\tilde{\alpha} \beta}$, by applying the inverse of the Gram matrix to (\ref{eq:uCPS_TDVP_equation}), the TDVP update for $C$ is obtained via (\ref{eq:dC_uCPS_TDVP}).

\noindent \emph{Efficient Implementation -} As discussed at the end of Section \ref{sec:CPS}, naively following an optimal uMPS  contraction ordering in the uCPS setting  will yield a computational cost that scales at best as $O(\log(D) D^3)$ for a local Hamiltonian, and we expect to be able to do better.  Details of a contraction ordering such that the cost of calculating the right hand side of (\ref{eq:uCPS_TDVP_equation}) scales as $O(D^3)$ are given in Appendix \ref{app:uCPS_contraction}. 
 
In addition to this, it is necessary to find an $O(D^3)$ implementation for the action by the inverse of the $D^2 \times D^2$ matrix $G$ on equation (\ref{eq:uCPS_TDVP_equation}).  Stemming from the fact that our parameterisation of variations $dC$ satisfying the tangent left gauge condition (\ref{eq:dC_uCPS_TDVP}) depends in a complicated way on the eigenvalue decomposition of the transfer matrix, the uCPS Gram-matrix turns out to be a much more complex object than its uMPS equivalent (given simply by $\rho_l \otimes \rho_r$ (\ref{uMPS_local_gram})), and is given by:
\begin{align}
\label{eq:uCPS_gram_matrix_explicit}
G^{ \tilde{\gamma} \delta | \tilde{\alpha} \beta} = \sum_{ij} \lambda_{ij} T^{ \tilde{\gamma} \delta | \tilde{\alpha} \beta}_{ i j} \ ,
\end{align}
where 
\begin{align}
T^{ \tilde{\gamma} \delta | \tilde{\alpha} \beta}_{ i j} = \left(  \conj{V}_{(R )\ i }^{\tilde{\gamma}} \conj{V}_{(L) \ j}^{\delta}   \right) (.*) \left(   V_{(R) \ i }^{\tilde{\alpha}} V_{(L) \ j }^{\beta}   \right) \ ,
 \end{align}
 and
 \begin{align}
 \lambda_{ij} = V_{(L) \ i }^{0} V_{(R) \ j }^{0}   (.*) ( 1 (. /) E_{  i j}) \ .
 \end{align}
Naively, both the cost of calculating $G^{-1}$, as well as its action on a vector, scales as $O(D^6)$; an obvious approach to reduce this cost would be to attempt to write  $G$ in a decomposition that would allow the action of its inverse on a vector to be calculated with cost $O(D^3)$, analogous to the manner in which a $O(d^2 D^6) \rightarrow O(d D^3)$ reduction in cost is achieved for uMPS TDVP. As far as we have been able to ascertain, this is not possible for (\ref{eq:uCPS_gram_matrix_explicit}).  While such an inverse \emph{does} exist for the matrix obtained by replacing all tilde-indices in (\ref{eq:uCPS_gram_matrix_explicit}) by their non-tilde extensions, and is given by inverting all the constituent matrices and taking $\lambda_{ij} \rightarrow 1 (./) \lambda_{ij}$, the truncation of the eigenvalue-one eigenvector seems to conclusively obstruct achieving any  appropriate decomposition for $G^{-1}$ itself. It is nevertheless possible to use an iterative method (such as the biconjugate gradient algorithm) in order to calculate the action  of $G^{-1}$ and retain $O(D^3)$ efficiency, since this only requires that the action of $G$ on an arbitrary vector be calculable with cost $O(D^3)$.  

 
A further caveat to the above is that in order to achieve $O(D^3)$ scaling, the number of iterations in the iterative subroutine required to achieve some desired accuracy must scale as a constant for large enough $D$, and this is not guaranteed.  For example, the number of iterations \emph{does} scale worse than constant  if the pre-conditioning step is not implemented appropriately (its implementation is  described in Appendix  \ref{app:uCPS_precond}). However, beyond this observation, for all the examples studied in this paper the iterative subroutine is observed to scale as $O(D^3)$.  
  
 Finally, it should be noted that while the general uMPS TDVP algorithm does not require an iterative subroutine for the Gram-matrix inverse step in order to achieve optimal efficiency, it  does require such a subroutine in order to calculate the third term in (\ref{eq:TDVP_rhs}) with cost $O(d D^3)$. The same term in uCPS TDVP, on the other hand, can be calculated explicitly with cost $O(D^3)$, as is described in Appendix \ref{app:uCPS_contraction}.  


\section{Properties of \MakeLowercase{u}CPS ground-state approximations}
\label{sec:properties_of_uCPS_gs}

In this section the imaginary time uCPS TDVP algorithm is used to obtain uCPS ground state approximations for a number of exemplary models: the quantum Ising model in a transverse magnetic field, the XY-model, and the Heisenberg model (all spin-$\frac{1}{2}$). The general aim is to study the capacity of uCPS  to capture ground state properties both at and away from criticality, while carefully considering the effects of uCPS basis-dependence.   We provide a detailed study of the convergence properties of uCPS ground state energy estimates, and of the convergence/scaling\footnote{ The entanglement entropy and the correlation length do not converge at criticality, so only the scaling properties of these quantities can be considered.} behaviour of the entanglement entropy and of the correlation length, all with respect to the size of the uCPS overlap $n$ (or, equivalently, with respect to the bond dimension $D$,  related to the overlap by $D = s^n$).   In addition,  the performance of the uCPS TDVP algorithm is analysed, and compared with the general uMPS imaginary time TDVP algorithm applied to the same models. 

\noindent \emph{Basis dependence -} In order to demonstrate the effect of uCPS basis choice, 
we calculate the uCPS groundstate energy estimates for the transverse field Ising Hamiltonian:
\begin{align}
\hatH = \sum_{i \in \mathbb{Z}} -J \hatsigma^z_i \hatsigma^z_{i+1} + h \hatsigma^x_i  \ ,
\label{eq:ising_hamiltonian}
\end{align}
the orientation of which remains fixed while the uCPS basis is rotated from the $x$- to the $z$-direction. Here $\{ \hatsigma^x, \hatsigma^y, \hatsigma^z \}$ are the Pauli matrices, $J$ determines the coupling strength between nearest neighbour spins, and $h$ determines the strength of the magnetic field; the model is critical for $\frac{h}{J} = \pm 1$.

The results for 2-site uCPS overlap ($D=4)$, at $h=1$, are displayed in the plot in Figure  \ref{fig:energy_vs_angle_overlap=2}.  At the points at which the two branches cross, imaginary time TDVP converges to a unique global minimum; this occurs for rotations away from the $z$-basis by integer multiples of $\frac{\pi}{2}$, i.e. for $z$- and $x$- basis choices, and at one additional intermediate angle, the precise value of which depends upon uCPS overlap. Away from the special points at which the branches cross in Figure \ref{fig:energy_vs_angle_overlap=2}, TDVP converges to the global minimum only for a certain fraction of TDVP runs initiated from a random uCPS state. 

The branching occurs due to the fact that, except for very special basis choices, uCPS breaks the two-fold degeneracy that is present in the uMPS approximation of the ground state at $h=1$.\footnote{It should be noted that the \emph{exact} ground state of the  quantum Ising model is doubly degenerate for $h<1$, but at finite bond dimension the $h=1$ uCPS/uMPS approximations are still effectively in the ferromagnetic phase (see e.g. \cite{2008PhRvB..78b4410T}, and also the discussion relating to Figure \ref{fig:Phase_transition_Ising_all}).} Thus, two states that would have exactly the same energy in the uMPS approximation acquire slightly different energies with uCPS due to the rotation dependence of the uCPS Ansatz. TDVP converges to one of these two solutions, depending upon details of the initial random state. At present, we do not understand precisely how the outcome is encoded in the initial state. A hint might perhaps be contained in the fact that a local unitary cannot be pushed through the copier tensor, so that the CPS parametrisation divides the Hilbert space into topologically distinct classes \cite{2015PhRvL.114q7204F}.    As one would expect, the branching is present for all values of $h$ in the ferromagnetic phase, and the two branches collapse as one enters the paramagnetic phase.


\begin{figure}[!htb]
\centering
\includegraphics[width=8cm]{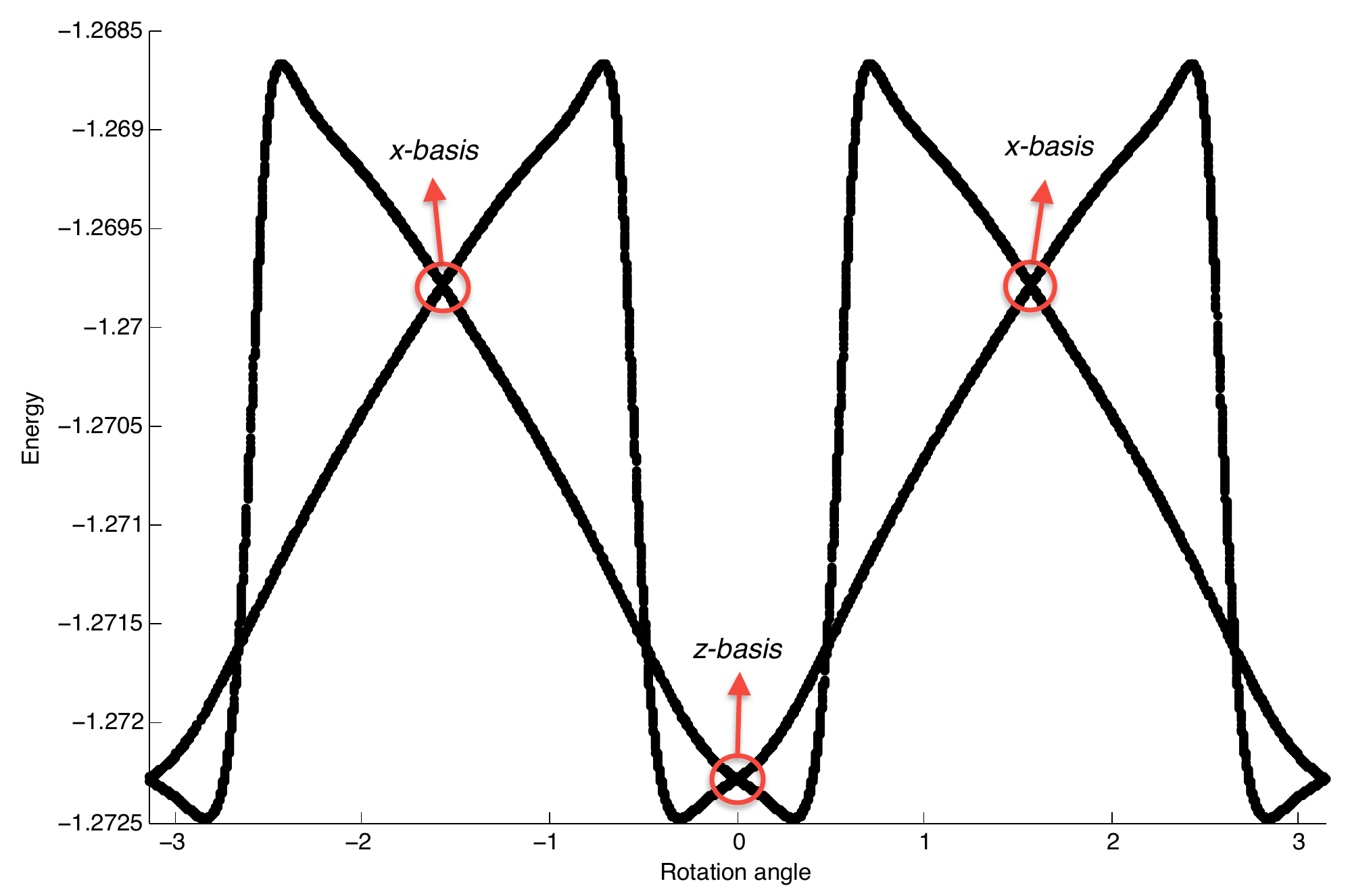}
\caption{\footnotesize{ (Colour online) The behaviour of 2-site overlap uCPS ground state energy approximations for the quantum Ising model at criticality,  as a function of the global rotation angle around the $y$-axis away from the $z$-basis choice. The TDVP algorithm exhibits unique convergence to the global minimum only at the points marked by the red circles, which denote the $z$- and $x$- basis choices, and at one intermediate point. In general, depending upon the details of the initial random state, imaginary time TDVP will converge to one of the minima associated with the splitting of otherwise degenerate energy levels, which occurs due to the lack of rotation invariance of the uCPS Ansatz.   } }\label{fig:energy_vs_angle_overlap=2}
\end{figure}

 In general, imposing constraints on uMPS can potentially introduce local minima on the variational manifold.   For example, uMPS TDVP often gets stuck in local minima when uMPS matrices are restricted to be real, and in that case some of the local minima can be singular.\footnote{Meaning that TDVP has in fact converged to a uMPS of a lower bond dimension than that of the initial random state. As some eigenvalues of the Gram-matrix go to zero in this limit, the TDVP algorithm becomes numerically unstable, at least in its naive implementation (for an implementation of MPS TDVP that avoids such instabilities see \cite{2014arXiv1408.5056H}).} In the present situation the local minima do \emph{not} correspond to singular points on the uCPS manifold, and thus no associated numerical complications are encountered. Moreover,  as long as the number of branches is constant as a function of uCPS overlap,  in order to be certain of having reached the ground state with some probability, the algorithm needs to be run a fixed number of times: $O(2)$ times for the example of the quantum Ising model in the ferromagnetic branch away from the special points. 
 



Let us next consider the XY-model, which is described by the Hamiltonian:
\begin{align}
\label{eq:XX_hamiltonian}
\hatH = -  \sum_{i \in \mathbb{Z}}  \frac{1 + \gamma}{2} \hatsigma^x_i \hatsigma^x_{i+1} + \frac{1 - \gamma}{2} \hatsigma^y_i \hatsigma^y_{i+1} + h \hatsigma^z   \ .
\end{align}
The ground state is two-fold degenerate for $0<h<1$ and $0 < \gamma < 1$. In this regime we observe two branches, as in the case of the quantum Ising model.  For $\gamma = 0$ the symmetry of the Hamiltonian is increased to $U(1)$, and here the number of local minima that uCPS converges to is not bounded as the size of the uCPS overlap is increased, as far as our numerical study is able to ascertain, and the $O(D^4) \rightarrow O(D^3)$ gains in the scaling of computational costs associated with the special structure of uCPS seem to be lost for this model.  In addition, no basis choice exists for which TDVP always converges to a global minimum.  

When the symmetry is increased to $SU(2)$, however,  as e.g. for the  spin-$1/2$  Heisenberg model, described by the Hamiltonian:
\begin{align}
\label{eq:heisenberg_hamiltonian}
\hatH = \sum_{i \in \mathbb{Z}} J \left( \hatsigma^x_i \hatsigma^x_{i+1} + \hatsigma^y_i \hatsigma^y_{i+1}  + \hatsigma^z_i \hatsigma^z_{i+1}  \right) \ ,
\end{align}
multiple branches are no longer observed. Full rotation invariance means that optimal uCPS approximations of physical quantities must be completely independent of the choice of basis, and moreover, in the case of the Heisenberg model, TDVP is always observed to converge to a global minimum.  

At this stage it can be observed that, in a loose sense, a judicious basis choice corresponds to one that is 'optimally' aligned with the entanglement generating terms in the Hamiltonian.  Clearly, for the quantum Ising model at criticality, the $z$-basis is maximally aligned with $\hatsigma^z \otimes \hatsigma^z$, and as demonstrated provides better energy estimates than those obtained by picking  the $x$-basis at equal overlap. Another desirable property with this choice is that TDVP always converges to the global minimum. However, as indicated by the plot in Figure \ref{fig:energy_vs_angle_overlap=2}, the best energy estimate is in fact achieved at an intermediate angle, but the computational downside of picking this point is that TDVP may get stuck in a local minimum, and that one needs to scan in order to find the orientation that achieves the lowest energy. In addition, the behaviour of uCPS is clearly highly model dependent. For example, the XY-model for $\gamma = 0$ has \emph{no} basis choice at which convergence is unique,  so scanning in this context may be a more sensible strategy than for the quantum Ising model.  Moreover, due to the $U(1)$ invariance of this model, the number of minima that TDVP converges to seems to increase indefinitely as overlap size is increased.  Since the TDVP algorithm converges to the global minimum only for a small fraction of runs, the algorithm becomes essentially statistical, and it is therefore questionable whether imaginary time TDVP has any advantages here over  Monte Carlo, the standard approach for CPS/String Bond type Ansaetze. For the $SU(2)$ invariant Heisenberg model the symmetry is large enough to eliminate such issues, but since all basis choices yield the same approximations for physical quantities, there is little motivation left for using a basis dependent Ansatz in the first place. In conclusion, what is meant by an 'optimal basis' is a function of both the model under investigation, and what it is that one wishes to achieve.  uCPS provides a good demonstration, in a well controlled context, why identifying a good basis is such a difficult problem  for general CPS/string bond states.    Nevertheless, attempting to maximally align the basis with entanglement generating terms certainly seems to be a good general guiding principle.


\noindent \emph{ Ground state convergence  at criticality -}  Next we investigate the ground state convergence properties of the transverse field quantum Ising model at criticality, comparing uCPS in the $z$- and  $x$- bases with uMPS. As noted, TDVP always converges to the global minimum for these basis choices.  The plot in Figure \ref{fig:energy_convergence_CPS_and_MPS_Ising} demonstrates that the uCPS ground state energy estimates are better for the $z$- than for the $x$-basis choice (that this is the case for 2-site uCPS overlap is already evident from Figure \ref{fig:energy_vs_angle_overlap=2}).   The more surprising observation is that, while the uMPS energy converges roughly polynomially as a function of bond dimension $D_{\mathrm{(uMPS)}}$ (since the model is critical), the uCPS energy converges instead polynomially \emph{with the size of the uCPS overlap} - note that the $x$-axis in the plot is shared between uCPS overlap size $n$ and $D_{\mathrm{(uMPS)}}$.  Thus, since $D_{\mathrm{(uCPS)}} = 2^n$, to achieve the same order of accuracy in the energy estimates uCPS requires an exponentially larger bond dimension than uMPS. Complementing this result is another surprising observation, depicted in the plot in Figure \ref{fig:uCPS_vs_uMPS_TDVP_convergence_times_Ising_h=1}: the total time needed to reach convergence to same accuracy scales in the same way for uMPS with respect to bond dimension as for uCPS with respect to overlap size. 
\begin{figure}[!htb]
\centering
\includegraphics[width=8cm]{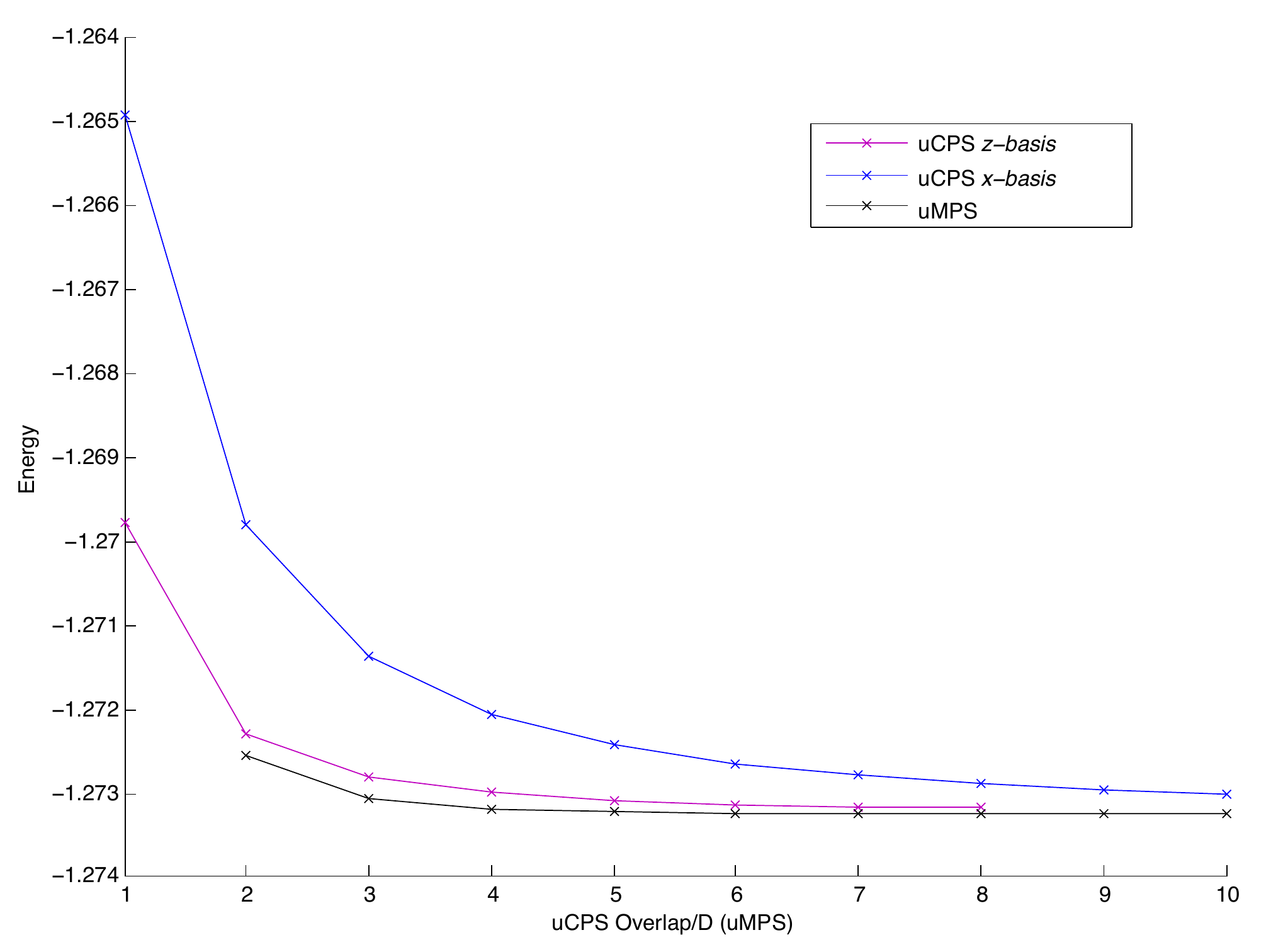}
\caption{\footnotesize{ (Colour online). Convergence of uCPS and uMPS ground state energies for the transverse Quantum Ising model at the criticality. The $x$-axis denotes overlap size for uCPS data, and bond dimension for uMPS data. } }\label{fig:energy_convergence_CPS_and_MPS_Ising}
\end{figure}

Thus, the indications from this example are that, even though the \emph{computer memory} costs are exponentially larger for uCPS than for uMPS,  the \emph{computer time} needed to reach some desired accuracy for the ground state energy approximation scales in the same way for both uCPS and uMPS.  This result seems to be generic, and is observed to hold for all the models studied in this paper.

\begin{figure}[!htb]
\centering
\includegraphics[width=8cm]{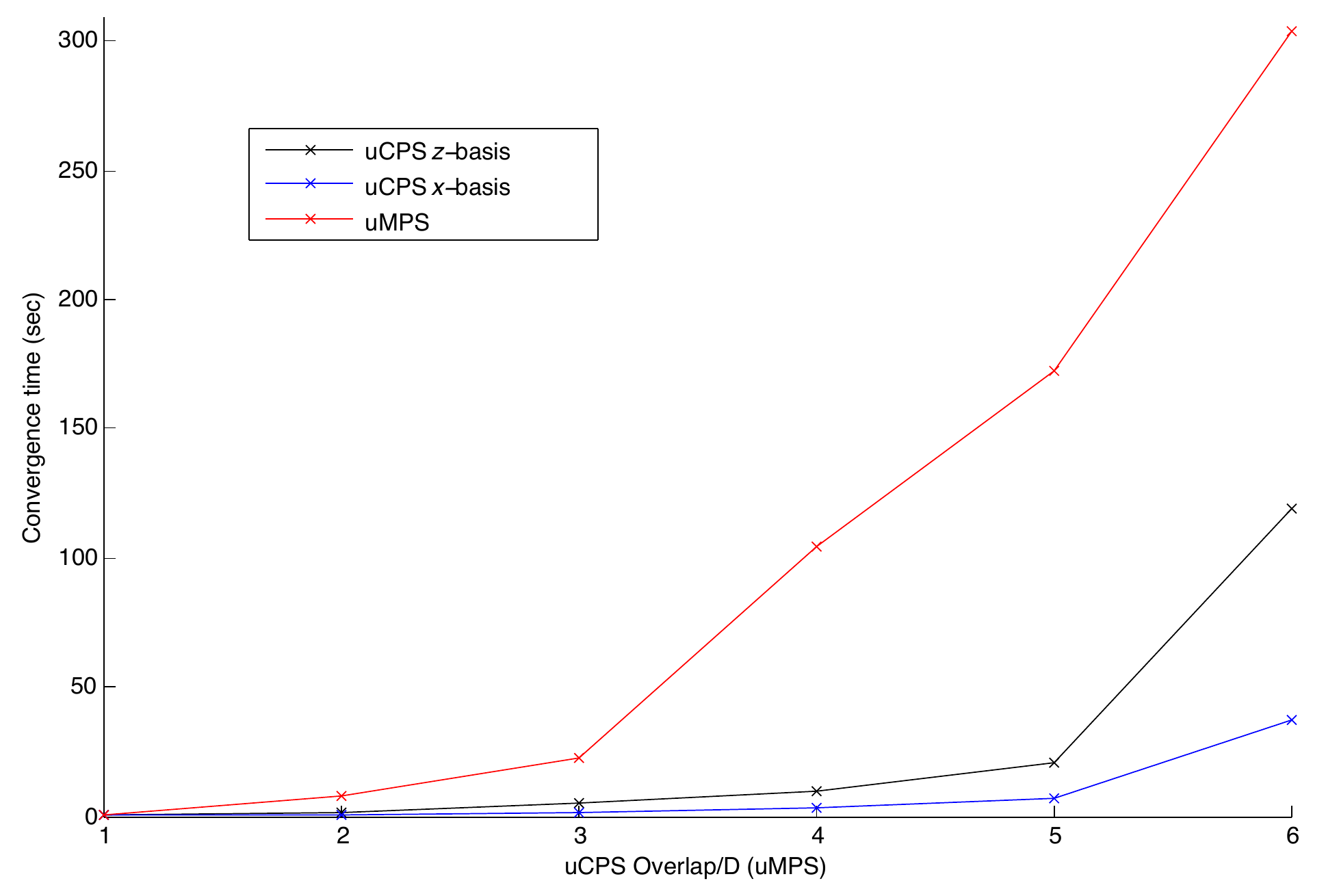}
\caption{\footnotesize{ (Colour online). Convergence times for uCPS and uMPS approximations for the ground state of the quantum Ising model at criticality.  The $x$-axis denotes overlap size for uCPS data, and bond dimension for uMPS data.  } }\label{fig:uCPS_vs_uMPS_TDVP_convergence_times_Ising_h=1}
\end{figure}

In this context it is also useful to consider the $SU(2)$ invariant Heisenberg model, so that it is impossible to pick out some preferred basis choice.  While one avoids the complications associated with basis dependence, the result of using an orientation dependent Ansatz on a rotation invariant Hamiltonian seems to manifest itself in extremely slow convergence, as a function of uCPS overlap, of the ground state energy approximations compared to what is observed for models whose entanglement generating terms point in a definite direction, such as e.g. the quantum Ising model.  This is illustrated by the plot in Figure \ref{fig:energy_convergence_CPS_Heisenberg}. In addition, the times required for TDVP to converge are much longer, at equal uCPS overlap, than for the quantum Ising model.  For example, the maximal bond dimension that we could achieve in reasonable time is $D_{\mathrm{uCPS}}=2^6$, and the corresponding energy density reproduces the correct result only to three digits. This slowdown originates in the iterative subroutine of the TDVP algorithm, which requires a much  larger number of iterations to converge to the required accuracy.
\begin{figure}[!htb]
\centering
\includegraphics[width=8cm]{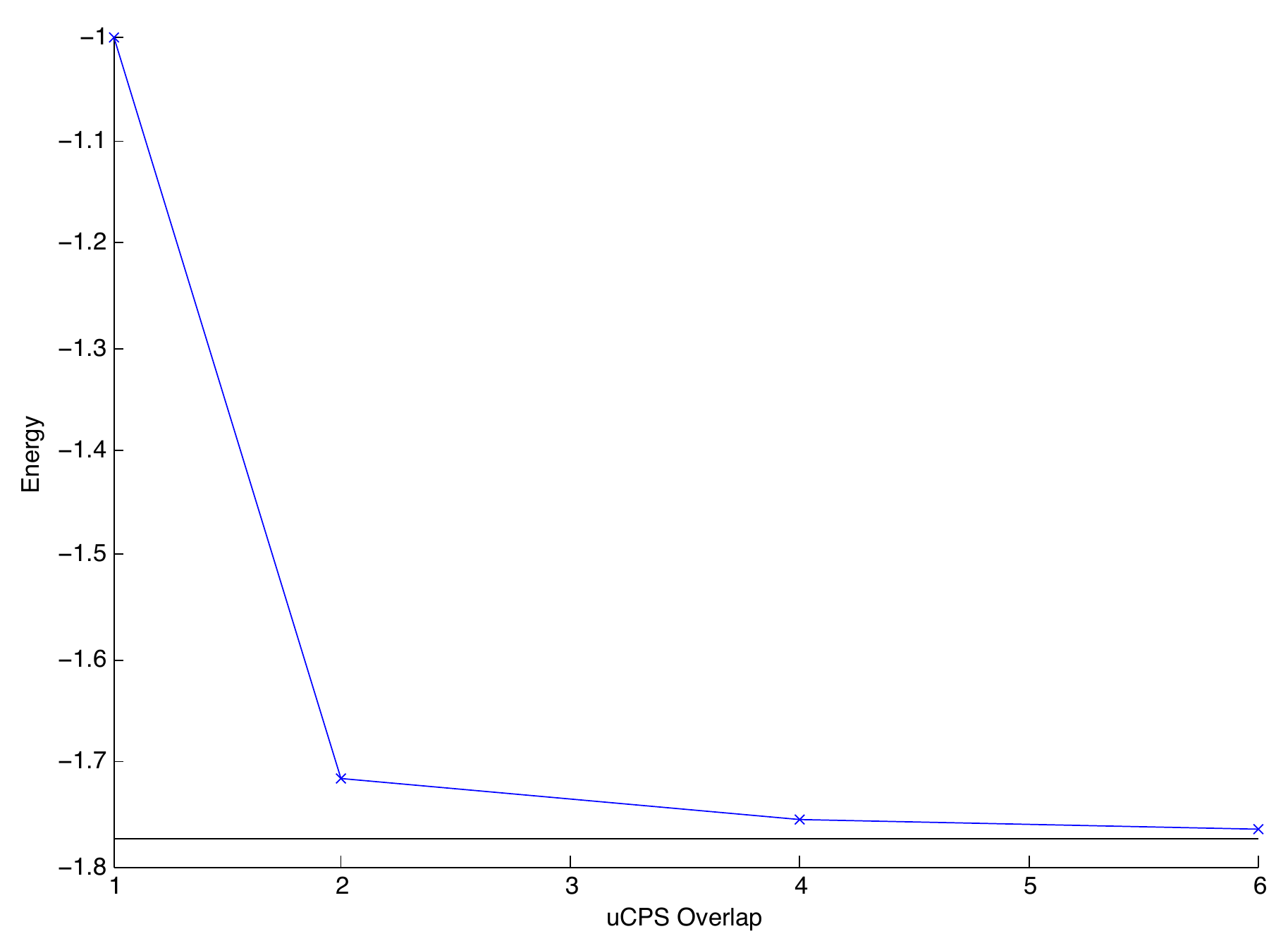}
\caption{\footnotesize{ (Colour online). Convergence of uCPS ground state energies for the Heisenberg model.  The black horizontal line denotes the exact energy. Whilst the uCPS approximations are not basis dependent, convergence turns out to be very slow.} }\label{fig:energy_convergence_CPS_Heisenberg}
\end{figure}

\noindent \emph{ Ground state convergence in a gapped phase -}  Next we examine the ground state convergence of uCPS  deep in the gapped phase.   The observations made above, regarding the computational time and memory costs of uCPS vs. that of uMPS, remain true away from criticality. Here we shall demonstrate that the analogous observations  also hold  for two rather different types of physical quantities, namely the \emph{correlation length} $\mu$ and the \emph{entanglement entropy} $S$. The correlation length is obtained from the largest eigenvalue $\lambda_2$ of the transfer matrix smaller than $1$ as:
\begin{align}
\label{eq:uCPS_corr_length}
\mu_{\mathrm{uMPS}} =  -\frac{1}{ \log( \lambda^{(\mathrm{uMPS})}_2)}  \ \ \  , \ \ \  \mu_{\mathrm{uCPS}}(n) = -\frac{n}{ \log( \lambda^{(uCPS)}_2)} \ ,
\end{align}
 for uMPS and uCPS, respectively, where $n$ is the uCPS overlap.  The entanglement entropy $S$ is the simplest measure, for pure quantum states defined on a region $\cR$, of entanglement between a sub-region $\cA \in \cR$ and the rest of the system. It is defined as:
\begin{align}
\label{eq:entanglement_entropy}
S =  - \mathrm{tr} ( \rho_{\cA} \log (\rho_{\cA} ))  = -  \sum_i \lambda_i^2 \log ( \lambda_i^2) \ ,
\end{align}
where  $\lambda_i$ are the Schmidt coefficients corresponding to the density matrix $\rho_{\cA}$ associated with the sub-region $\cA$. One can show that the Schmidt coefficients corresponding to a cut of the infinite spin chain into two semi infinite sub-chains are given by the singular values of $\rho_l^{\frac{1}{2}} \rho_r^{\frac{1}{2}}$, where $\rho_l$ and $\rho_r$ are the left and right uMPS environment matrices defined in Appendix \ref{app:MPS_review}; for uCPS left and right environment eigenvectors see (\ref{eq:uCPS_left_eigv}) and (\ref{eq:uCPS_right_eigv}). 

The convergence of $S$ for a half-infinite chain, for the quantum Ising model  in the paramagnetic phase (specifically for $J=1$, $h=0.5$), is depicted in the plot in Figure \ref{fig:entropy_vs_overlap_Ising_h=0.5}. We note that for the $z$-basis choice convergence of $S$ approaches that of uMPS. This is not surprising, given that in the limit $h\rightarrow0$ the ground state approaches a product state aligned along the $z$-basis, and so one would expect uCPS in this basis to be capable of capturing the exact state accurately already at small overlap. 

\begin{figure}[!htb]
\centering
\includegraphics[width=8cm]{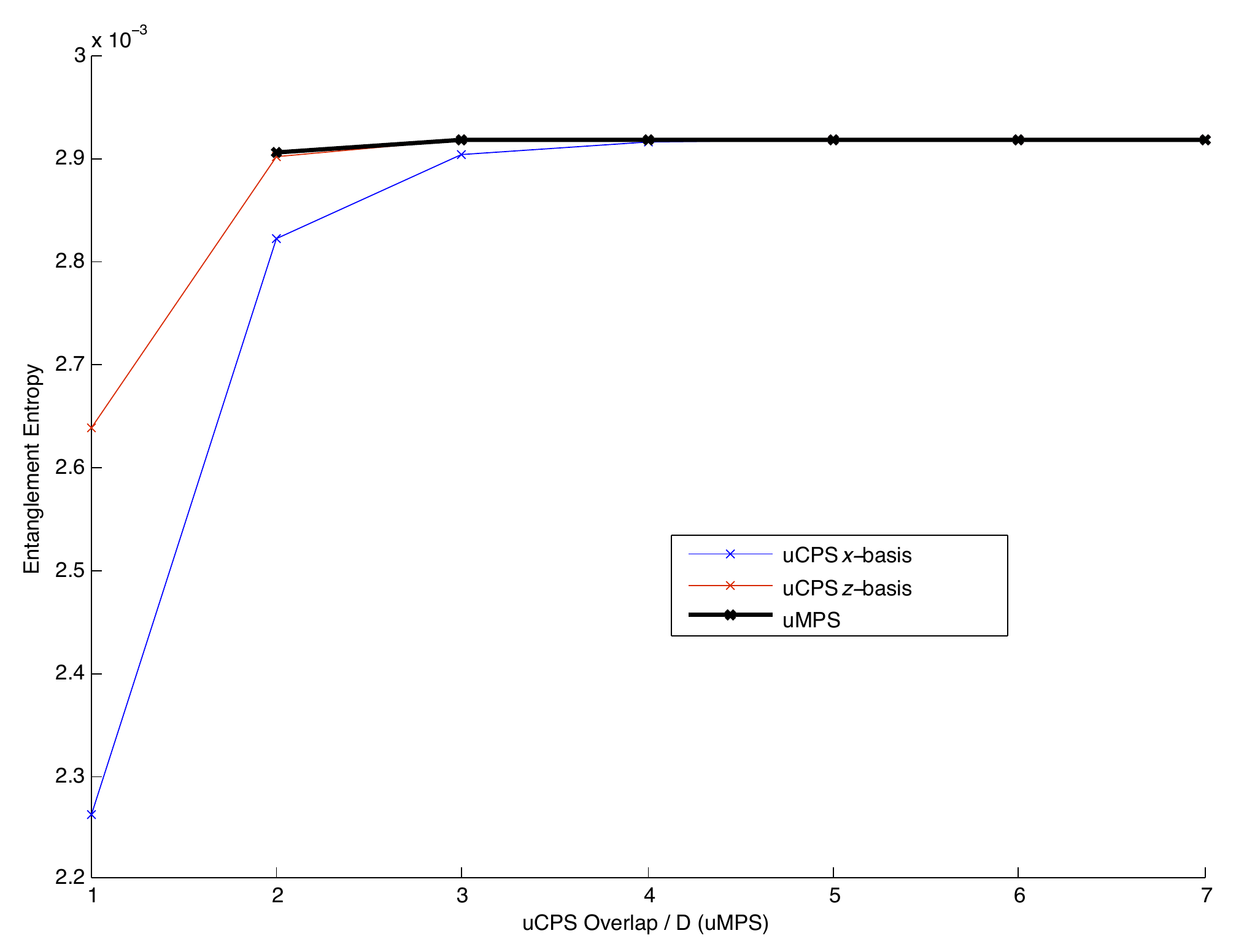}
\caption{\footnotesize{ (Colour online). Entanglement entropy $S$  as function of uCPS overlap/uMPS bond dimension $D$, for the quantum Ising model in the ordered gapped phase ($J=1$, $h=0.5$).  }}\label{fig:entropy_vs_overlap_Ising_h=0.5}
\end{figure}


The present example also demonstrates that uCPS is able to reproduce ground state energies to machine precision  both for $z$- and $x$-basis choices. Due to the exponentially higher memory cost of uCPS compared to uMPS, in practice it is necessary to be far in the gapped phase to observe such convergence;  for example, for $J=1$, $h=0.5$, machine precision is achieved for both basis choices with a 7-site uCPS overlap (which already corresponds to $D_{\mathrm{uCPS}} = 128$), while for  $J=1$, $h=0.7$, the necessary overlap size is out of reach for a desktop with 16 GB of RAM. Remarkably, a naive first-order implementation of uCPS TDVP is sufficient to achieve this; despite the fact that most of the Gram matrix eigenvalues are zero to machine precision in this regime, no numerical instabilities are encountered.  This behaviour depends crucially on implementing an appropriate pre-conditioner  in the iterative subroutine step  responsible for applying the inverse of the Gram matrix to a vector, as is described in Appendix \ref{app:uCPS_precond}.


 
The convergence of the correlation length with uMPS bond dimension/uCPS overlap is depicted in the plot in Figure \ref{fig:corr_length_vs_overlap_Ising_h=0_5}. A notable feature is that, for both $x$- and $z$-basis choices, uCPS convergence is much smoother than for uMPS. Such behaviour is also observed at criticality, as is  discussed later in this section (see Figure \ref{fig:uCPS_entropy_vs_log_mu}).
 
\begin{figure}[!htb]
\centering
\includegraphics[width=8cm]{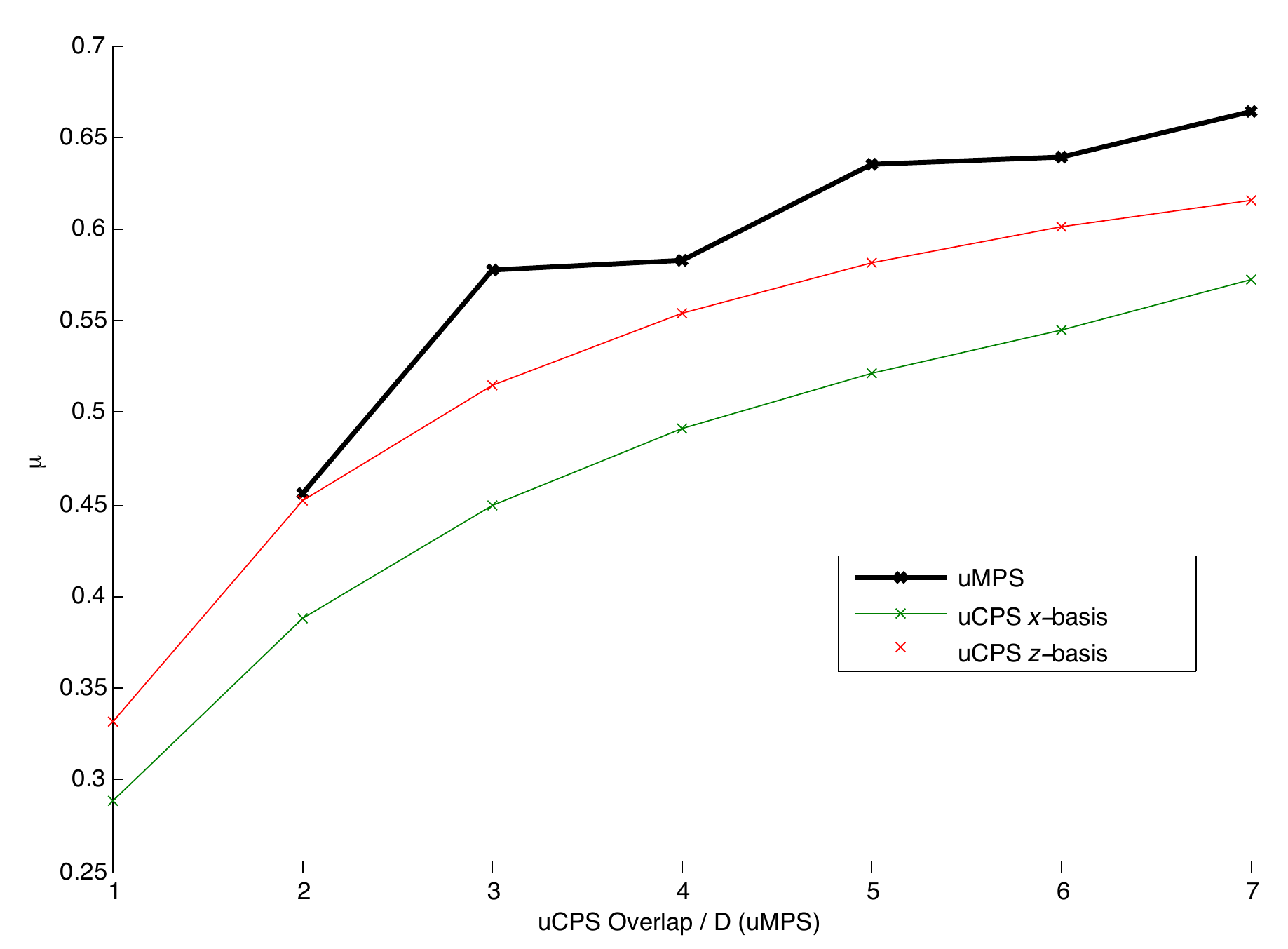}
\caption{\footnotesize{ (Colour online).  Correlation length  $\mu$ as function of uCPS overlap/uMPS bond dimension $D$, for the quantum Ising model in the ordered gapped phase ($J=1$,  $h=0.5$). }}\label{fig:corr_length_vs_overlap_Ising_h=0_5}
\end{figure}

Finally let us also note that tor the XY model, both at and away from criticality,  convergence properties are similar to those of the quantum Ising model for $\gamma$ close to one; as $\gamma$ approaches zero convergence becomes slower, and the TDVP algorithm more sensitive to integration errors.
 
\noindent \emph{Finite entanglement scaling at criticality - }  Next we study the scaling properties of the correlation length  (\ref{eq:uCPS_corr_length}) and entanglement entropy (\ref{eq:entanglement_entropy}) for uCPS at criticality. Since the area law no longer holds, these quantities, while finite at any given bond dimension $D$, grow indefinitely as $D \rightarrow \infty$.  Since a finite $D$ bounds the amount of entanglement that can be encoded in a CPS/MPS,  scaling with respect to $D$ is referred to as \emph{finite entanglement scaling}, and has been extensively studied in the uMPS setting \cite{2008PhRvB..78b4410T, 2009PhRvL.102y5701P, 2012PhRvB..86g5117P, 2015PhRvB..91c5120S}. In what follows,  we demonstrate that also at criticality overlap size plays the same role for uCPS as the bond dimension does for uMPS, and furthermore show how universal quantities can be calculated using uCPS finite entanglement scaling.

Let us first consider the scaling of the correlation length $\mu$ (\ref{eq:uCPS_corr_length}) with uCPS overlap. The work \cite{2008PhRvB..78b4410T}  provides numerical evidence that for a critical system uMPS  scales with bond dimension as $( D_{\mathrm{uMPS}})^\kappa$ in the limit of large $D_{\mathrm{uMPS}}$, where $\kappa$  is a universal constant. In \cite{2009PhRvL.102y5701P} it was furthermore argued that $\kappa$ only depends upon the central charge $c$ of the critical system via the relation:
\begin{align}
\label{eq:kappa_c_relation}
\kappa = \frac{6}{c \left( \sqrt{ \frac{12}{c}} + 1 \right)} \ .
\end{align}
The plot in Figure \ref{fig:uCPS_Ising_h=1_log_mu_vs_log_N} demonstrates that, to high accuracy, for uCPS  $\mu$ is instead proportional to $n^{\tilde{\kappa}}$, where $n$ is the uCPS overlap, and  $\tilde{\kappa}$ a constant.   For the critical quantum Ising model,  a linear fit of $\log(\mu)$ vs. $\log(n)$  for uCPS in the $z$-basis yields  $\tilde{\kappa} =1.004 \pm 0.006 $. It seems plausible that the exact value of $\tilde{\kappa}$ is equal to one in the limit $n\rightarrow \infty$, i.e. that the correlation length is exactly proportional to uCPS overlap size. In the $x$-basis, as can be seen in Figure \ref{fig:uCPS_Ising_h=1_log_mu_vs_log_N},  a significant oscillation over the whole range of available $n$ is present,  and making any conclusion regarding the  value of $\tilde{\kappa}$ in the limit $n \rightarrow \infty$ is difficult. Based upon the available data it is thus not possible to say whether or not $\tilde{\kappa}$ is basis dependent in this limit. This in turn makes it impossible to make any assertion as to whether the value of $\tilde{\kappa}$ is universal (and if so, in what sense).
 \begin{figure}[!htb]
\centering
\includegraphics[width=8cm]{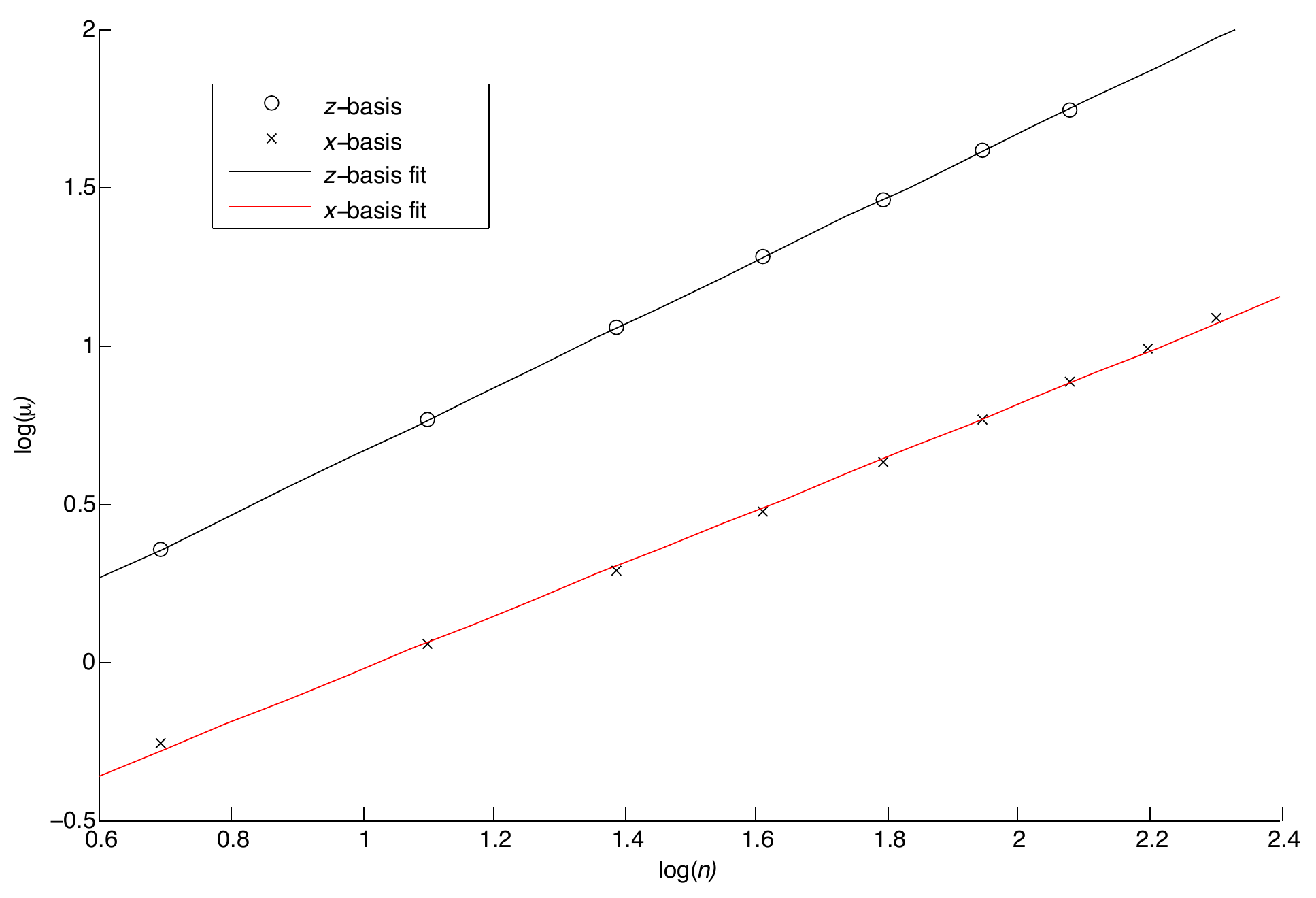}
\caption{\footnotesize{ (Colour online). The logarithm of the  correlation length $\mu$ versus the logarithm of uCPS overlap for uCPS ground state approximations for the quantum Ising model at criticality, for both the $x$- and $z$- basis choices. In the $z$-basis the linear fit is particularly accurate, with slope equal to one to good approximation.  } }\label{fig:uCPS_Ising_h=1_log_mu_vs_log_N}
\end{figure}

 
At present we can therefore not see any manner in which, in the uCPS context, something akin to  (\ref{eq:kappa_c_relation}) could be used to estimate the central charge $c$. Nevertheless, uCPS can be used to calculate universal quantities along the lines of  \cite{2012PhRvB..86g5117P, 2015PhRvB..91c5120S}, where  it has been demonstrated that a particularly powerful way to calculate critical exponents and the central charge is to scale not directly with respect to $D$, but with respect to the uMPS correlation length $\mu(D)$.  The central charge, for example, can be estimated from the scaling of the entanglement entropy (\ref{eq:entropy_vs_logmu_inf})  with the correlation length (\ref{eq:uCPS_corr_length}) as follows. For a  $(1+1)$ critical system  it has been shown \cite{1994NuPhB.424..443H, 1742-5468-2004-06-P06002} that the entanglement  entropy corresponding to an interval $\cA$ of length $x_{\cA}$  grows as:
\begin{align}
\label{eq:entropy_vs_logmu_open}
S_{\cA} =   \frac{c}{3} \log (x_{\cA}) + k \ ,
\end{align} 
where $c$ is the central charge of the system, and $k$ a constant.   The entanglement entropy of the half-infinite line then scales as:  
\begin{align}
\label{eq:entropy_vs_logmu_inf}
S =   \frac{c}{6} \log (\mu) + \tilde{k} \ ,
\end{align} 
where $\mu$ is some length scale introduced in the system - in our case this is precisely the correlation length associated with finite bond dimension (\ref{eq:uCPS_corr_length})  - and $\tilde{k}$ is again some constant.  Relation (\ref{eq:entropy_vs_logmu_inf}) in conjunction with (\ref{eq:uCPS_corr_length}) can thus be used to obtain an estimate of the central charge. 


\begin{figure}[!htb]
\centering
\includegraphics[width=8cm]{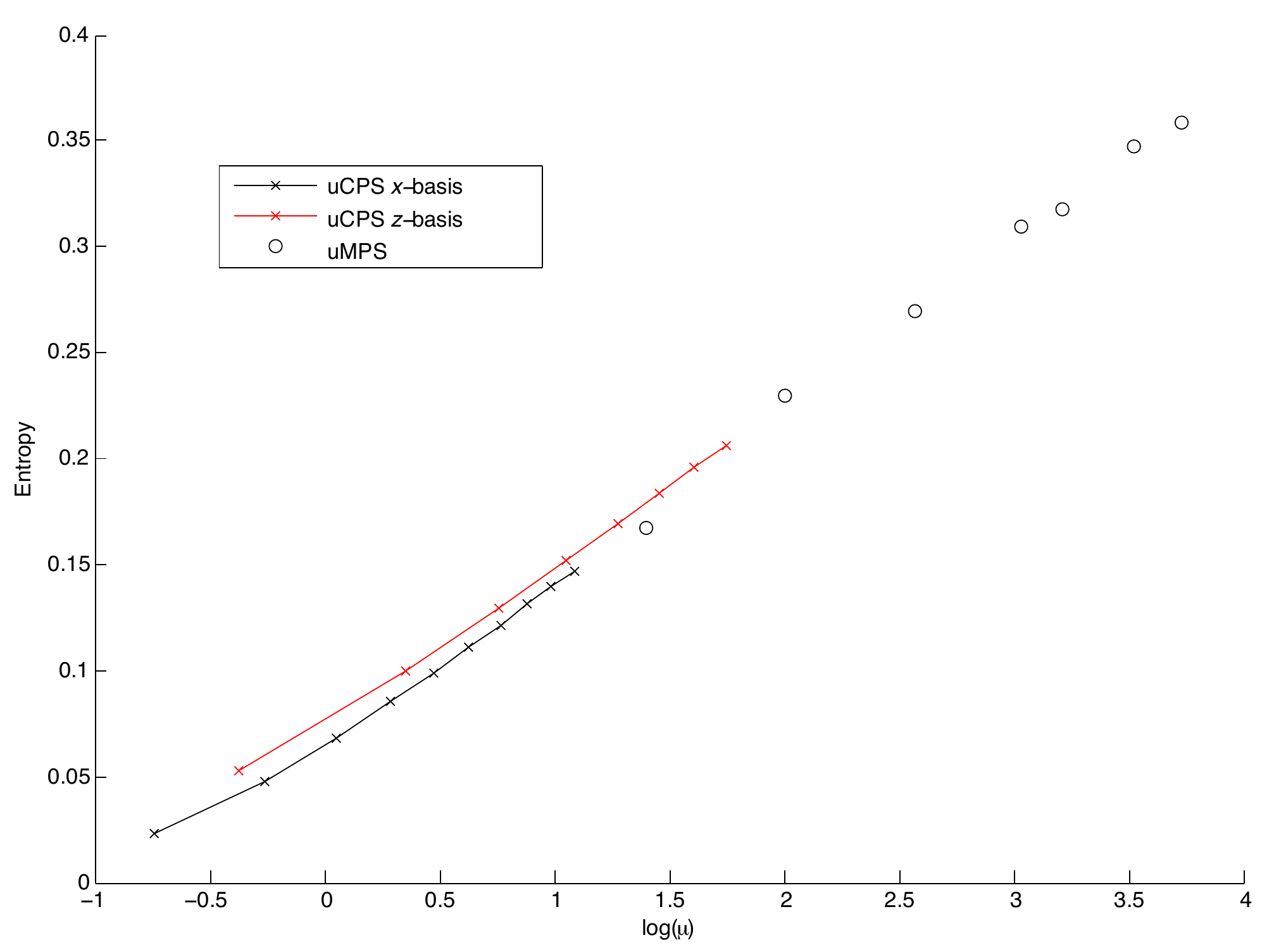}
\caption{\footnotesize{ (Colour online). The scaling of uCPS/uMPS entanglement entropy $S$ vs $\log(\mu)$, where $\mu$ is the correlation length. The slope approaches a constant value, which is theoretically predicted to be $\frac{c}{6}$, where $c$ is the central charge. } }\label{fig:uCPS_entropy_vs_log_mu}
\end{figure}

  The scaling of $S$ vs. $\log(\mu(n))$ for the quantum Ising model at criticality is depicted in the plot in Figure  \ref{fig:uCPS_entropy_vs_log_mu}. As one can see, oscillations in the entanglement entropy as a function of $\log (\mu)$ observed for uMPS are absent from uCPS in a fixed basis. Performing a simple  $1/n$, $n \rightarrow \infty$ extrapolation for the slope of this curve, with  $z$-basis data up to $n=10$, yields the estimate $c = 0.50142...$ for the central charge. This deviates from the exact value of $c=\frac{1}{2}$ only in the third digit, and provides better accuracy than the result obtained via finite uMPS entanglement scaling using all bond dimensions from $D=2$ to $D=64$ in \cite{2015PhRvB..91c5120S}. The time needed to generate this uMPS data on an average spec current desktop, even when using more advanced techniques than the simple first-order implementation of imaginary time TDVP (such as the conjugated gradients method or iDMRG),  is of the order of a week. On a comparably powerful computer the uCPS data used here was generated in a few hours.  It should be noted, however, that uMPS finite entanglement methods are much more accurate for critical exponent than for the central charge estimates, and that the exponentially larger memory cost of uCPS means that uMPS is capable of accessing states with a lot more entanglement (as Figure  \ref{fig:uCPS_entropy_vs_log_mu} clearly demonstrates). Nevertheless, for the range of bond dimensions for which uCPS has a sufficient amount of RAM, the lack of oscillations means that uCPS can in practice be useful for  making accurate estimates much more quickly than is possible with uMPS. 
 
It is also interesting that uMPS and uCPS data points in Figure  \ref{fig:uCPS_entropy_vs_log_mu}  lie roughly on the same line, and while uMPS data is significantly noisier, it seems to be bounded by the optimal $z$- and the suboptimal $x$- uCPS basis choices.  This gives an indication that there may be some relationship between the oscillations in the uMPS data and the rotation invariance of the uMPS Ansatz.

\noindent \emph{In conclusion -} While  a single TDVP step scales better with bond dimension for uCPS than for uMPS,  an exponentially larger amount of computer memory is needed in order to achieve the same accuracy with uCPS than with uMPS. However,  the computational time required to obtain the same accuracy, surprisingly, scales in the same manner for both. The precise nature of the scalings is  sensitively dependent upon both the model under investigation and the choice of basis. For models with degenerate ground states,  convergence to the global minimum is in general only achieved for a certain fraction of TDVP runs. The uCSP algorithm exhibits certain advantages over uMPS when calculating universal quantities using finite entanglement scaling; oscillations found in uMPS are not present for uCPS, so scaling can be deduced more accurately with comparable computer time (but larger computer memory) cost. This seems to occur precisely {\it because} of the fixing of the basis.  It would be particularly useful if oscillations are eliminated for CPS/string bond states in a similar manner beyond one dimension, where obtaining large number of points for a range of bond dimensions, as may be necessary with oscillations present, can be prohibitively expensive.

\section{Quenches with \MakeLowercase{u}CPS}
\label{sec:CPS_quenches}

In this section we use real time TDVP applied to uCPS to study quenches across the critical point in the quantum Ising model (\ref{eq:ising_hamiltonian}) which exhibit so-called \emph{dynamical phase transitions}.  Dynamical phase transitions can occur whenever the return amplitude, 
\begin{align}
\label{eq:return_amlitude}
G(t) =  \bra{\Psi_0} e^{-iHt} \ket{\Psi_0}  \ ,
\end{align}
also referred to as the Loschmidt amplitude, has zeros \cite{PhysRevLett.101.120603}.  The rate function for the return probability (referred to from here on just as the \emph{rate function}):
\begin{align}
\label{eq:rate_funciton}
l(t) = - \lim_{L \rightarrow \infty} \frac{1}{L} \log  | G(t) |^2   \ ,
\end{align}
can acquire non-analyticities, analogous to those present in the free energy in a thermodynamic setting. For a uCPS/uMPS, this quantity corresponds simply to the logarithm of the second largest eigenvalue of the transfer matrix. 

The thermodynamic equivalent of (\ref{eq:return_amlitude}) is obtained from purely imaginary time evolution, i.e. $t= -i \tau$ with $\tau$ real. In this case $G(\tau)$ corresponds to the canonical partition function of a system with finite length in the $\tau$ direction, and with boundaries described by $\ket{\Psi_0}$.  \emph{Equilibrium phase transitions} of the system are in correspondence with the zeros of $G(\tau)$ that occur as one takes the thermodynamic limit $L \rightarrow \infty$, which, if present, result non-analyticities in the free-energy $- \log | G(\tau) |^2$. It should be noted that, while the thermodynamic partition function can acquire zeros only in the thermodynamic limit, for the return amplitude  (\ref{eq:return_amlitude})  this can also happen also at finite system size. Thus, in general no simple correspondence between equilibrium and dynamical phase transitions exists \cite{2014PhRvB..89l5120A}.  

The rate function (\ref{eq:rate_funciton}) can be calculated exactly for the quantum Ising model, and a dynamical phase transition occurs for quenches across the critical point \cite{2013PhRvL.110m5704H, 2013PhRvB..87s5104K}, so in this setting a simple correspondence between the dynamical and equilibrium cases does exists.   The ground state of the quantum Ising model (\ref{eq:ising_hamiltonian}) undergoes an equilibrium phase transition from a ferromagnetic phase for $h<1$ to a paramagnetic phase for $h> 1$. This transition is captured accurately by uMPS even at small bond dimension, with the expectation value of order parameter operator  $ <  \hatsigma^z >$ 
going from a positive value  for $h < h_c$, to zero\footnote{To machine precision.} for $h > h_c$, with $h_c$ approaching the exact value $h_c=1$ from above with increasing bond dimension.   The uCPS behaviour is highly basis-dependent, as is demonstrated by the plot in Figure \ref{fig:Phase_transition_Ising_all}. The critical point is approximated with comparable accuracy  by uMPS and uCPS in the $z$-basis, at bond dimension equal to uCPS overlap, but the uCPS approximation gets increasingly less accurate for choices of basis away from $z$,
and completely misses the phase transition in the $x$-basis. 

\begin{figure}[!htb]
\centering
\includegraphics[width=8cm]{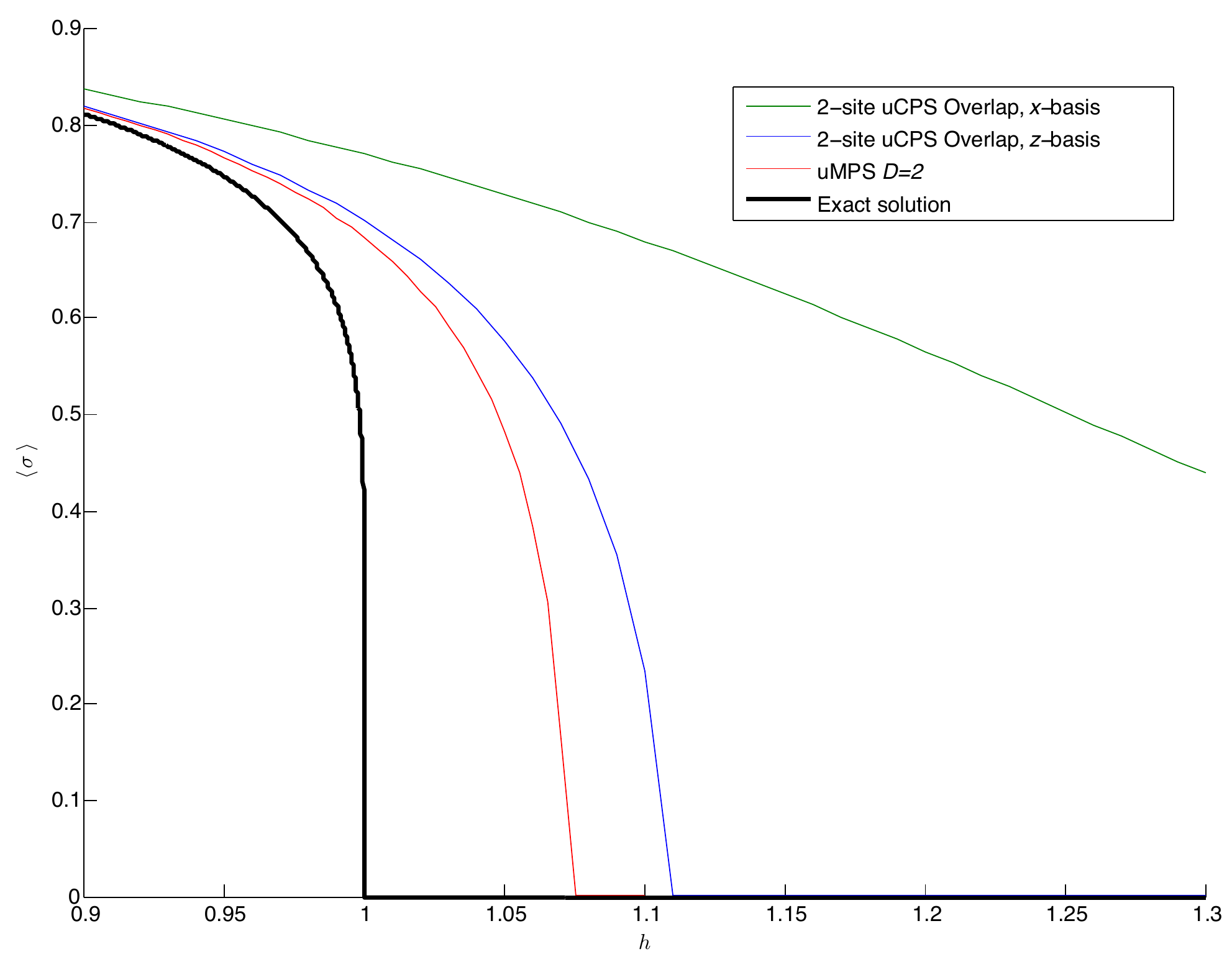}
\caption{\footnotesize{ (Colour online). Both uCPS and uMPS are capable of capturing the quantum Ising model equilibrium phase transition, from the ferromagnetic phase, $h<1$, characterised by a non-zero value for the order parameter operator expectation value $< \sigma^z >$, to the paramagnetic phase $h>1$ when $< \sigma^z > = 0$.  The accuracy of the uCPS approximation for the critical point, $h=1$, degrades as a function of the rotation angle away from the $z$-basis. In the $x$-basis the phase transition is entirely missed. }}\label{fig:Phase_transition_Ising_all}
\end{figure}

In the following, we concentrate on a quench initiated in the paramagnetic phase, with $h=1.5$ in the quantum Ising Hamiltonian (\ref{eq:ising_hamiltonian}), with the time evolution performed according to a Hamiltonian with the magnetic field (instantaneously) changed to $h=0.1$; the post-quench Hamiltonian is therefore deep in the ferromagnetic phase. The reversed quench, from the ferromagnetic to the paramagnetic phase, also exhibits a dynamical phase transition, but is more unwieldy to analyse \cite{2013PhRvB..87s5104K}. As it yields very similar conclusions regarding the properties of uCPS, it will not be explicitly discussed here.

\begin{figure}[!htb]
\centering
\includegraphics[width=8cm]{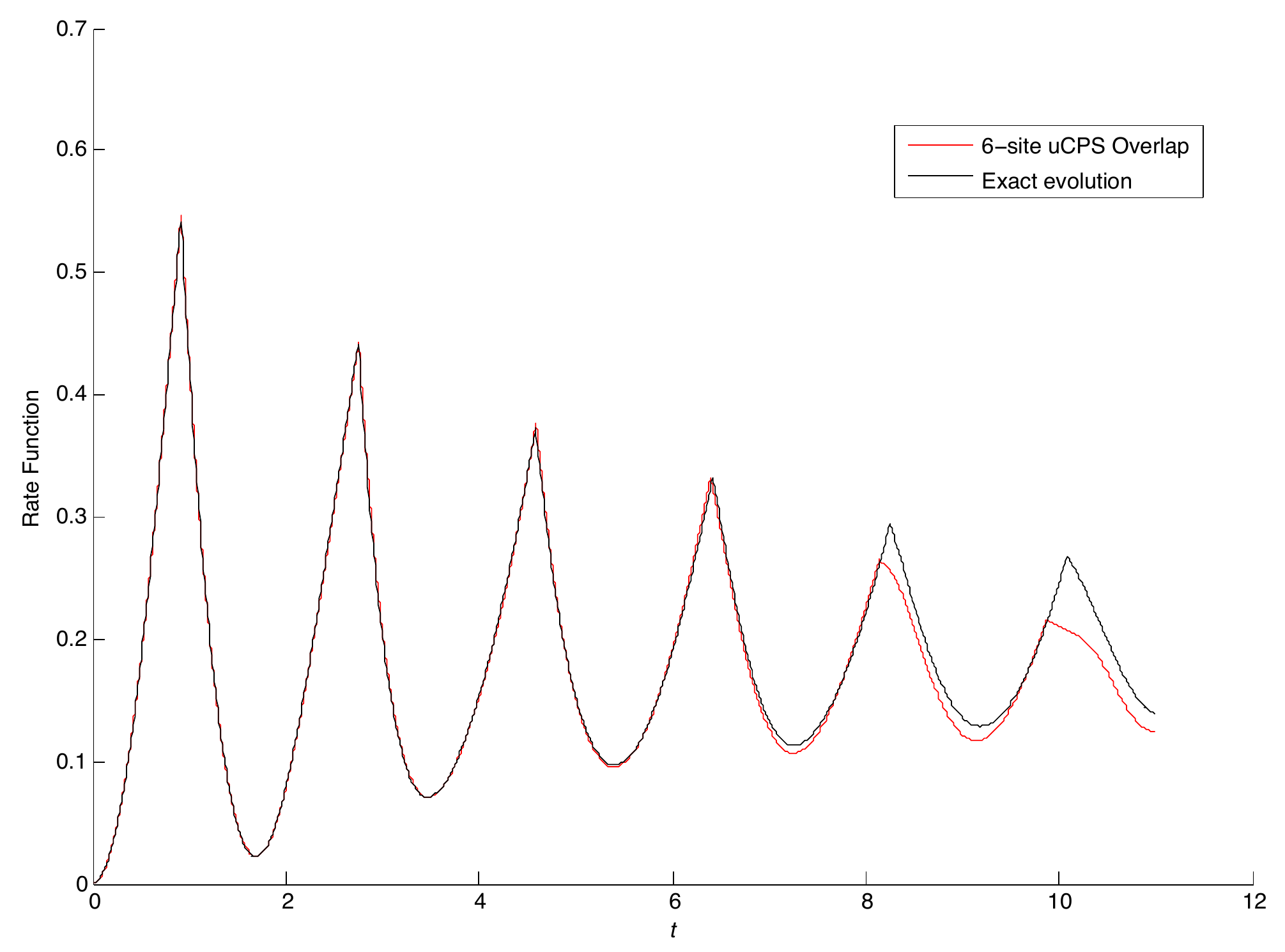}
\caption{\footnotesize{ (Colour online). The rate function vs. time in the $z$-basis for 6-site uCPS overlap compared with the exact evolution. The quench corresponds to the ground state of the quantum Ising Hamiltonian with $h=1.5$ (paramagnetic phase) evolved with the $h=0.1$ Hamiltonian (ferromagnetic phase).  } }\label{fig:rate_function_vs_time_Overlap=6_zzx}
\end{figure}

In the  $z$-basis uCPS captures the dynamical phase transition very accurately, as demonstrated in the plot in Figure  \ref{fig:rate_function_vs_time_Overlap=6_zzx}. In addition to capturing non-analyticities, simulations with uCPS in this basis exhibit approximate recurrences at large times, i.e. beyond the point at which the exact rate function is captured correctly, for all values of uCPS overlap. For single-site overlap recurrences are in fact exact.  Plots of uCPS approximations for single- and 3-site overlap in the $z$-basis   are depicted in the two top plots of Figure \ref{fig:rate_function_vs_time_recurrence} and clearly show recurrences. 

In contrast, uCPS in the $x$-basis (Figure \ref{fig:rate_function_vs_time_Overlap=2_5_xxz}) do not capture the non-analyticities in the rate function at all. For times prior to the first non-analyticity, the accuracy of the uCPS approximation of the rate function increases with increasing overlap, however the non-analyticity is never actually captured. Beyond the point at which the non-analyticity  occurs in the exact function, the behaviour of the uCPS approximation completely misses the correct behaviour and is chaotic, not converging to any definite function with increasing uCPS overlap size. It should be noted that non-analytic behaviour is not observed at any time in the $x$-basis uCPS approximation. This is very different to the large time behaviour of uCPS in the $z$-basis. 


The behaviour of the uMPS approximation of the quench, at large times, exhibits a combination of that observed for $z$- and $x$-basis uCPS approximations in the following sense: as can be seen in lowermost plot in Figure \ref{fig:rate_function_vs_time_recurrence}, beyond the point at which the exact rate function is captured accurately (for a given bond order), the uMPS behaviour is chaotic, as is the case for uCPS in the $x$-basis, but it does exhibit non-analyticities at  arbitrarily large times, a feature observed for uCPS in the $z$-basis.  

\begin{figure}[!htb]
\centering
\includegraphics[width=8cm]{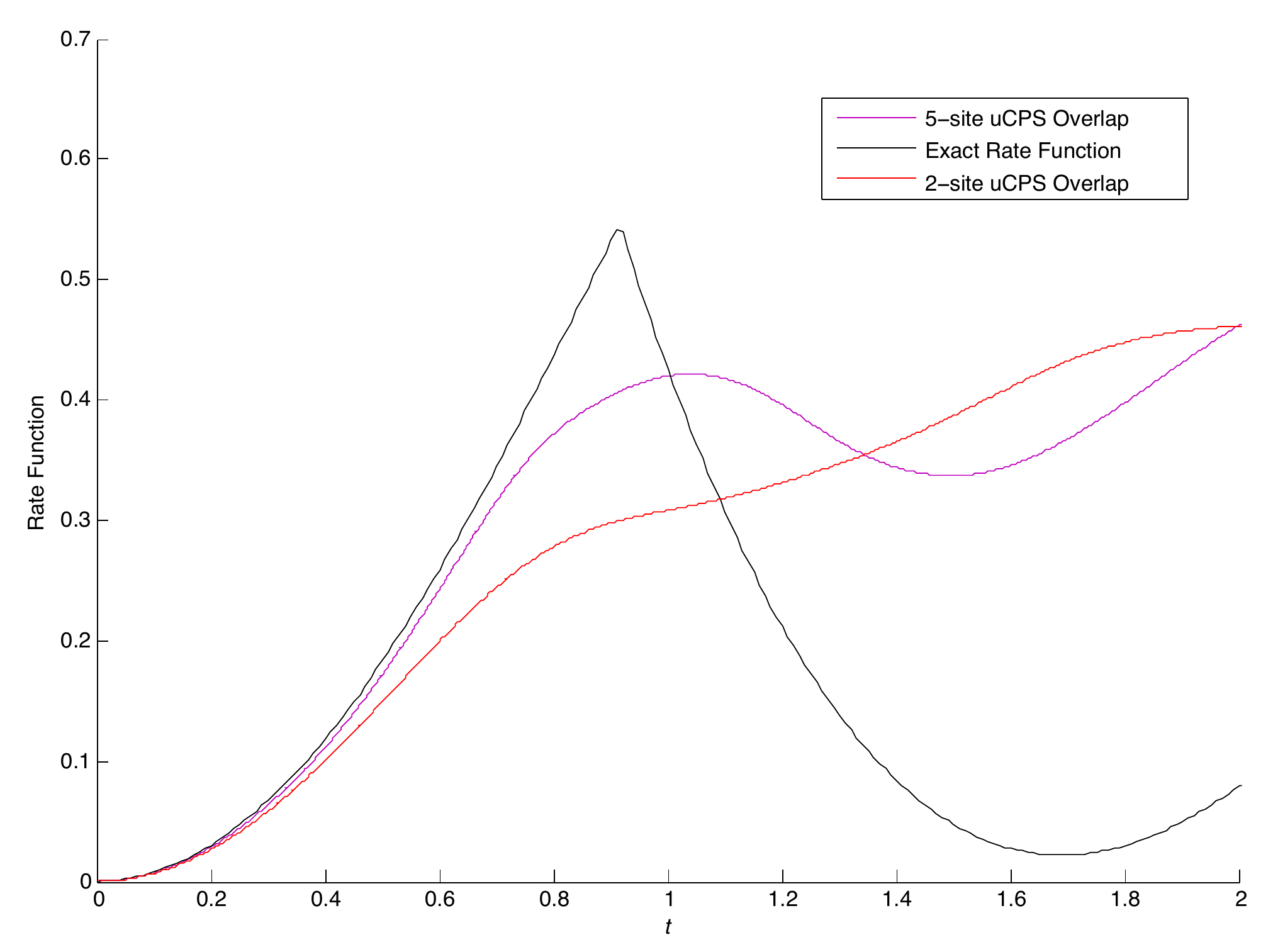}
\caption{\footnotesize{ (Colour online). The uCPS approximation, for 2- and 5-site uCPS overlap in the $x$-basis, of the rate function for the $h=1.5 \rightarrow h=0.1$ quench in the quantum Ising model.  In this basis uCPS completely misses the dynamical phase transition. The behaviour of the uCPS approximation is analytical even at large times (not displayed).} }\label{fig:rate_function_vs_time_Overlap=2_5_xxz}
\end{figure}

Comparing the $D=6$ uMPS approximation in Figure \ref{fig:rate_function_vs_time_recurrence} with the 6-site uCPS overlap approximation in the $z$-basis in Figure \ref{fig:rate_function_vs_time_Overlap=6_zzx} demonstrates that the  quality of the uCPS and uMPS approximations is comparable. This is another example - here in the context of real time evolution - of the general observation that the capacity of uCPS and uMPS to capture the properties of spin systems is roughly the same, in an optimal basis, for uCPS overlap equal to the uMPS bond dimension. In addition, we observe that, for the examples studied in this section, the time needed to run uMPS and uCPS approximations of the same quench at uCPS overlap size equal to uMPS bond dimension is roughly the same. This demonstrates that also in the context of quenches it is only the computer memory cost, and not the computer time cost, that scales exponentially worse for uCPS.

\begin{figure}[!htb]
\centering
\includegraphics[width=9cm]{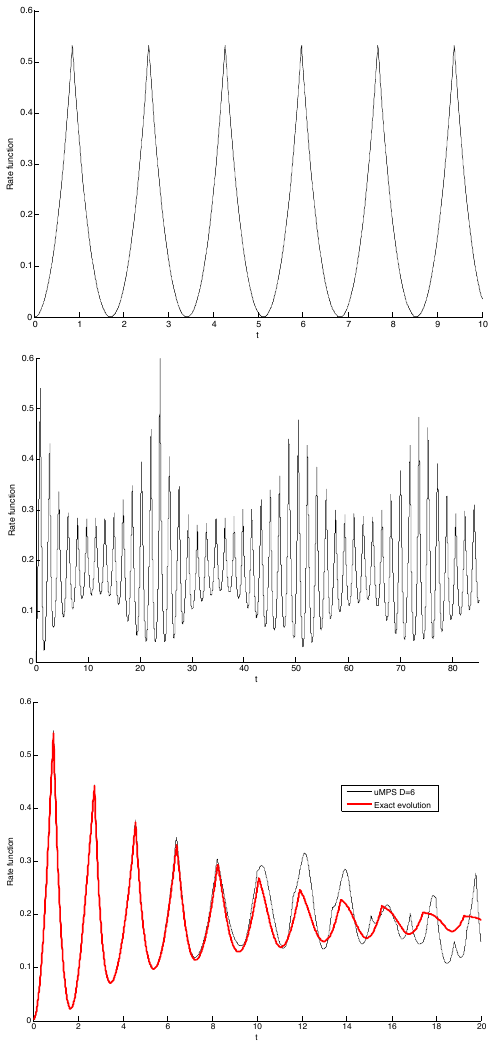}
\caption{\footnotesize{ (Colour online). Rate function vs. time in $z$-basis exhibits recurrence when projected to the uCPS manifold. Recurrence is exact for 1-site CPS overlap (top), and approximate when the overlap is larger than one, as demonstrated for the 3-site uCPS overlap case (middle).  The uMPS approximation does not exhibit recurrences at large times; beyond the point at which the rate function is accurately captured, the behaviour is chaotic (bottom), and the precise behaviour is also highly bond-dimension  dependent. } }\label{fig:rate_function_vs_time_recurrence}
\end{figure}

\section{Discussion}
\label{sec:discussion}

This paper has centred around developing the deterministic TDVP algorithm to study a restricted class of uCPS states. Such an algorithm is itself a departure as CPS states are usually optimised stochastically and - as far as we are aware - not used to study real time Hamiltonian evolution. Our analysis has revealed several interesting features of the numerical cost of using uCPS and their sensitivity to choice of basis.

\noindent \emph{ Computational cost of uCPS TDVP algorithm -} implementation of the TDVP algorithm shows interesting efficiency gains, aspects of which may be extendible to higher dimensions.  In Section \ref{sec:properties_of_uCPS_gs} it has been demonstrated that the cost of a single uCPS TDVP step scales, optimally, as $O(D^3)$. Since uCPS maps to an uMPS with bond dimension equal to the physical spin dimension, and the cost of one uMPS TDVP step is $O(d D^3)$, one naively expect the cost of each uCPS TDVP step to scale as $O(D^4)$, but the special structure of uCPS enables this to be improved upon. This special structure is most effectively analysed by considering the uCPS $\rightarrow$ uMPS mapping, which necessitates the introduction of the so called copier tensor, and the fact that a singe $n$-site coper factorises into a product of $n$ single-site copiers  already implies the reduction to $O( \log(D) D^3)$. The mapping of a generic CPS to PEPS involves copiers that factorise in a similar manner, so it is likely that an increase in efficiency is possible also in more than one dimensions for deterministic algorithms that respect the CPS structure.  

However, the efficiency with which a uCPS represents a state to a given degree of accuracy is not as great as one might naively anticipate. As discussed in Section \ref{sec:CPS}, we have considered a restricted set of CPS that are equivalent to MPS with bond and local Hilbert space dimension equal. One might expect then that uCPS with an optimally chosen basis are of comparable accuracy to uMPS of this bond order. This is not the case.  Both deep in the gapped phase and also at criticality, quantities of interest, such as the ground state energy, converge with  size of the uCPS overlap $n$ in roughly the same manner as they converge with bond dimensions for uMPS.  The implication is that, as far as physical observables are concerned, uCPS scaling is exponentially worse than uMPS, since $D_{\mathrm{uCPS}} = s^n$, where $s$ is the dimension of the spin of a single site. For critical systems  one can make a particularly precise statement: while for uMPS the  correlation length $\mu$ increases with bond dimension $D$ as $\mu \propto D^\kappa$, in the limit $D \rightarrow \infty$, for uCPS the correlation length scales instead as  $\mu \propto n^{\tilde{\kappa}}$, where $\kappa$ and $\tilde{\kappa}$ are constants.   It is not possible to determine, from the accessible range of uCPS overlap sizes, whether or not $\tilde{\kappa}$ is basis dependent, and thus, whether or not it encodes universal behaviour. Interestingly for the quantum Ising model in the $z$-basis $\tilde{\kappa}$ seems to be one to good accuracy, and so the correlation length is precisely proportional to the uCPS overlap.

Surprisingly, while the \emph{memory} cost is exponentially worse for uCPS than for uMPS,  the computer \emph{time} needed to fully converge to the optimal uCPS ground state approximation, using imaginary time TDVP initiated from a random state, scales in the same way for  uCPS with respect to overlap, as it does for uMPS with respect to bond dimension. Combined with the above result, that uCPS achieves roughly the same accuracy as uMPS for overlap equal to uMPS bond dimension,  this observation implies that uCPS in the optimal basis matches uMPS accuracy with the same computer time cost, but an exponentially worse memory cost. This does not contradict the statement that the cost of a single step in the imaginary time TDVP is exponentially worse for uCPS, both in time and memory cost (since at each step exponentially larger matrices need to be multiplied). The speedup observed with uCPS, over the course of the whole imaginary time TDVP run, reflects the fact that for uCPS much larger time steps can be taken than for uMPS before instability sets in. In addition, while for imaginary time TDVP integration errors due to excessively large time steps can conspire to actually aid convergence, both for uCPS and uMPS  - and of course as long as these time steps are not too large - this effect aids convergence more efficiently in the case of uCPS.  The net result is that the the exponentially higher cost of a single step for uCPS in the TDVP algorithm is counteracted by these two effects to bring the computer time cost of the whole TDVP run down, so that it actually scales exponentially better than would be expected when considering the costs of a single step of the algorithm; clearly, none of this improves the memory requirements.  

The computational costs are observed to have the same properties in the context of real-time TDVP, which has been used to simulate quantum quenches in Section \ref{sec:CPS_quenches}. The only caveat is that in the analysis, in place of considering the time necessary for imaginary time TDVP to converge,  one must compare the uCPS vs. uMPS cost of simulating the evolution of a quench over some reasonably long time. Given that integration errors are not of great concern for imaginary time TDVP, only a simple Euler-step integrator was used to generate the ground state approximations, and one may object that some of the conclusions made above regarding the cost of imaginary time TDVP are merely artefacts of the particularly simple type of integration scheme.  This is not the case, as can be checked by employing a more sophisticated  integration algorithm. Moreover,  our real time uCPS quench simulations were computed using  the adaptive step Runge-Kutta-Fehlberg 4(5) method, with the same conclusions regarding uCPS vs. uMPS computational costs.


\noindent \emph{ Basis dependence -} In addition to the above, basis dependence has a major impact on the behaviour of the uCPS TDVP algorithm. In Section \ref{sec:CPS_quenches} it was demonstrated that with a sub-optimal choice of basis important physics can be missed, as demonstrated by the fact that both dynamical and static phase transitions in the quantum Ising model are not observed if one chooses to work in the $x$-basis. Clearly, this is a cautionary lesson when working with higher dimensional CPS/string bond states.

In the imaginary time context, it is observed that the uCPS  TDVP algorithm converges to the global minimum irrespective of the choice of the initial random state in general only when the ground state is not degenerate. For degenerate vacua  the algorithm converges to the global minimum only for a certain fraction of runs initiated from a random state, except possibly for very special choices of basis for which convergence is unique.  If there is a finite number of degenerate vacua, the number of possible minima is in general equal to the degeneracy and does not increase with bond dimension. For a continuous symmetry group uCPS TDVP seems to converge - as far as our numerical analysis is capable of ascertaining - to a number of local minima that increases without bound with uCPS overlap. The statements concerning the scaling and cost of obtaining a ground state made above still hold,  but when convergence is not unique the algorithm acquires a probabilistic ingredient, and for continuous symmetries the scaling of the computational cost can increase by more than a constant factor.     

 In reference to models examined in Section \ref{sec:properties_of_uCPS_gs}, for the quantum Ising Hamiltonian entanglement is generated by the single term $\hatsigma^z \otimes \hatsigma^z$, and a number of suitable basis choices exist that yield unique convergence; for the XY model with $\gamma = 0$, which has $U(1)$ symmetry, no such choice exists.   At the other extreme convergence is observed to be unique, irrespective of basis choice, for the Heisenberg model, which is described by a rotation invariant Hamiltonian.  Here all basis choices are clearly equivalent, and imaginary time TDVP always converges to a unique minimum. However, energy converges extremely slowly compared to the other models studied (see Figure \ref{fig:energy_convergence_CPS_Heisenberg}) which demonstrates that uCPS is best suited to the study of Hamiltonians maximally aligned with the uCPS basis.

In conclusion, a good basis choice is one for which  the entanglement generating terms are 'optimally aligned', in some sense,  with the uCPS basis.   The problem of what precisely is meant by 'optimal' is encountered for all CPS/string bond state approaches, and is a difficult one to tackle with any generality - not only due to  technical challenges, but also because the answer depends  upon what precisely one wishes to achieve.  We have  illustrated how this pans out in detail in the context of uCPS in Section \ref{sec:properties_of_uCPS_gs}. In this case, convergence properties, probabilistic aspects of the algorithm, and  the accuracy of estimates for physical observables are all basis dependent, yet can not in general be optimised simultaneously; the choice depends upon which of these properties one wishes to prioritise.

\section{Conclusions}
\label{sec:conclusions}
Although quantum states in dimensions higher than one can be represented efficiently by tensor networks, physical properties may not  in general be calculated efficiently without further approximations. One way around this is to place additional restrictions upon the tensor network so that its properties are easier to calculate. Such restrictions inevitably involve compromises and a balance between efficiency gains and accuracy. We have investigated this balance in the controlled context of a restricted class of uniform one-dimensional Correlator Product States that may also be considered a restriction upon uniform matrix product states.  Similar restrictions may be applied in higher dimensions - Correlator Product States with small, double overlaps can be mapped to small bond order PEPS.

Our main results are:
\begin{itemize}
\item The application of the time dependent variational principle to uCPS. Usually CPS - as well as the more general class of string bond states - are optimised using a stochastic Monte-Carlo type approach. They are well-suited to this because of their efficient sampleability.  By considering the mapping of uCPS into uMPS, one naively expects a single uCPS TDVP step to scale as $O(D^4)$. 
Utilising the special structure uCPS, we have shown that the cost of a single uCPS TDVP step can be reduced to $O(D^3)$. Since this is based upon the properties of the copier tensor whose properties generalise to higher dimensional CPS $\rightarrow$ PEPS mappings, our analysis indicates that a similar reduction should be possible for higher dimensional deterministic algorithms that respect the CPS structure.   
\item The capacity of uCPS to capture physical information about a system scales exponentially worse with bond dimension than uMPS. In order to capture  the ground state energy, or a quantum quench, to the same accuracy as a $D$ dimensional uMPS, one has to work with uCPS with overlap size of the order $n \approx D$, which is exponentially more expensive since $D_{\mathrm{uCPS}} = s^n$, where $s$ is the dimension of a single spin. This is, surprisingly, only reflected in computer memory usage, not in the computer time needed to obtain the ground state via imaginary time TDVP, or to run quantum quenches, which for a uCPS with overlap of size $n$ is of the same order as for uMPS with $D=n$.  
\item The choice of uCPS basis has a strong effect upon the behaviour of the TDVP algorithm, on the capacity of uCPS to accurately approximate the physical system under consideration, as well as on the behaviour of uCPS under scaling. A good basis choice, generally speaking, has the property that it is closely aligned with the entanglement generating terms in the Hamiltonian.  With an optimal choice of basis uCPS will generally capture the physics as well as uMPS, for uCPS overlap equal to the uMPS bond dimension. On the other hand, with a suboptimal choice uCPS can completely fail to capture important physics such as equilibrium or dynamical phase transitions.   We also observe that, being an Ansatz that is not rotation invariant, except for special basis choices uCPS will in general break any degeneracy present in the exact ground state (or in the related uMPS approximation): depending on details of the random initial state, imaginary time TDVP will converge to local minima associated with this separation of otherwise degenerate energy levels.
\item  Having fixed a basis, the scaling of a physical quantity with bond dimension is much smoother for uCPS than for uMPS. In particular, the scaling of quantities such as entanglement entropy or the correlation length often has strong oscillations at lower bond dimensions in the case of uMPS, and these almost entirely disappear for uCPS in a fixed basis. 
\item Some properties of uMPS exhibit a combination of features that can be isolated by making judicious choices of uCPS basis.  For example, the oscillations seen in the scaling behaviour of uMPS  seem to be bounded by the smooth behaviour of uCPS scalings, at one end by the uCPS in the optimal  basis and the other by the least optimal choice of basis (see Figure \ref{fig:uCPS_entropy_vs_log_mu}). Similarly the behaviour of uMPS at large times for the quench exhibiting dynamical phase transitions  is a combination of the recurring non-analyticities seen for the optimal uCPS choice of basis, and the chaotic but analytic behaviour observed in the least optimal basis (see Figures \ref{fig:rate_function_vs_time_Overlap=2_5_xxz} and \ref{fig:rate_function_vs_time_recurrence}).
\end{itemize}

The analysis in this paper has mostly been geared towards bettering our theoretical understanding uCPS compared to standard uMPS, with a view to identifying characteristic properties of the uCPS variational manifold that may be of use when studying CPS/string bond states in general, and in particular in higher dimensions.  It should be stressed that  uCPS has potential practical advantages  already in  one dimension.   A general observation made at various points in this paper is that the  uCPS TDVP algorithm, applied to a suitable Hamiltonian and in an optimal basis, is very robust - both in its imaginary and real time variants (see e.g.  Figures \ref{fig:entropy_vs_overlap_Ising_h=0.5} and \ref{fig:rate_function_vs_time_recurrence}) -  under the right circumstances more so than a comparably costly uMPS TDVP run. A further advantage of uCPS is described at the end of Section \ref{sec:properties_of_uCPS_gs},  where it is shown that one can utilise the superior scaling properties of uCPS compared to uMPS in order to generate estimates of universal quantities in critical theories, with accuracies not achievable with comparable computational time cost using uMPS.  Finally, an aspect of our analysis  that has not been emphasised in the course of this paper is that, while uCPS yields similarly accurate estimates of physical quantities for overlap sizes equal to uMPS bond dimension, the actual bond dimensions accessed by uCPS are exponentially larger than what is accessible with uMPS at comparable computational time cost. For example, for the critical quantum Ising model the bond dimension $2^{10} = 1024$  uCPS ground state estimate in the  $x$-basis is reached with roughly the same computer time cost needed to generate the uMPS $D=10$ ground state approximation (see Figure \ref{fig:energy_convergence_CPS_and_MPS_Ising}). One interesting question is, for example, whether the uCPS $\rightarrow$ uMPS mapping described in Section \ref{sec:CPS} could  provide a more efficient way of initialising  a $D=1024$ iDMRG run, than e.g. by building it up from a $D=10$ uMPS state?




\begin{acknowledgements}
We gratefully acknowledge support by EPSRC under grant number EP/I031014/1 and EP/I004831/2.
T. \DJ uri\'c  acknowledges support from the EU Grant QUIC (H2020-FETPROACT-2014, Grant No. 641122). V. S.  would like to thank Damian Draxler, Jutho Haegeman,  Micha{\"e}l {Mari{\"e}n}, and Frank Verstraete for helpful discussions. 
\end{acknowledgements}

\appendix

\section{Review of Matrix Product States and the Time-Dependent Variational Principle}
\label{app:MPS_review}

A matrix product state (MPS) takes the form:
\begin{align}
\label{eq:MPS_def_app}
& \ket{ \Psi [ A ] }= \\ \nonumber & \sum_{i_1, i_2, \cdots, i_N}^d  v_L^\dagger A^{i_1}_1 A^{i_2}_2 \cdots A^{i_N}_N  v_R \ket{ i_1, i_2, \cdots, i_N }  \ ,
\end{align}
where $d$  is the number of physical (spin) degrees of freedom, and $A^{i_a}$ is a $D_{A-1}\times D_{A}$ matrix, while $v_L$ and $v_R$ are vectors of dimensionality $D_0$ and $D_N$ respectively. The dimension of these internal/virtual indices  is referred to as the \emph{bond dimension}.  The open boundary condition Ansatz is presented here for the sake of concreteness - the periodic boundary condition MPS Ansatz  corresponds to taking a trace in place of contraction by $v_L^{\dagger}$ and $v_R$.  Graphically the state (\ref{eq:MPS_def_app}) can be represented as: \newline 
\centerline{  \includegraphics[scale=0.3]{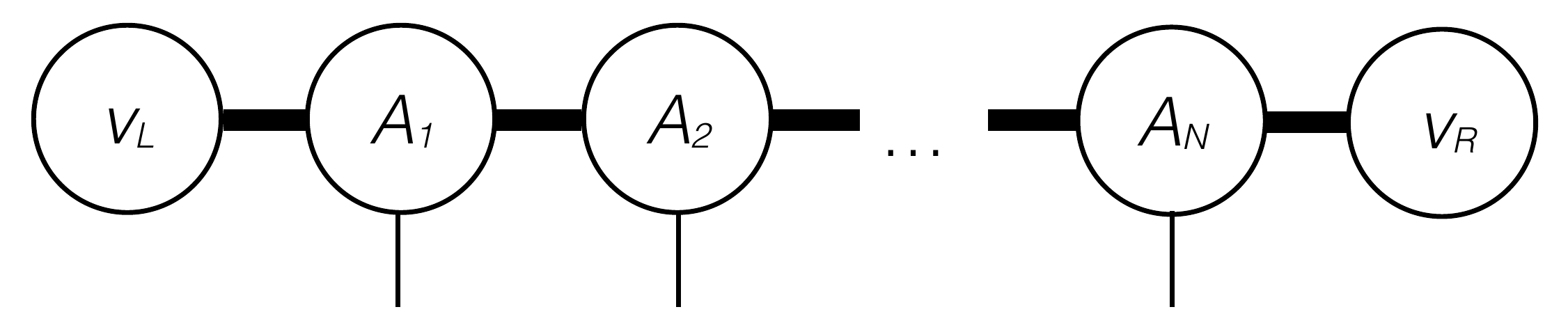}   \ \ \raisebox{+2.5ex}{,} } 
\newline where the thick lines represent the internal bond dimension indices, while the physical spins are denoted by the thin lines with 'uncontracted' ends.  Here and in what follows we will not graphically depict any variation in the dimensionality of physical or bond indices.

Transformations of the MPS  matrices of the form:
\begin{align}
\label{eq:gauge_transf}
A^{i_n} \rightarrow G_{n-1} A^{i_n} G_n^{-1} \ ,
\end{align}
where $G_n, G_{n-1} \in GL(D, \bC)$, leave the state invariant, and are referred to as gauge transformations. The gauge transformations of the boundary (co)-vectors are given by $v_L^\dagger \rightarrow v_L^\dagger G_0^{-1}$, and $v_R \rightarrow  G_N v_R$.  
 
Applying the time-dependent variational principle (TDVP;  see introduction to Section \ref{sec:uCPS_TDVP}) to the MPS variational class \cite{PhysRevLett.107.070601, 2013PhRvB..88g5133H}  yields:
\begin{align}
\dot{A}^*(t) = \argmin_{\dot{A}(t)} \norm{   \ket{\Phi(\dot{A}, A)} +  i \hatH (t) \ket{\Psi  (A(t)) }   } \ ,
\end{align}
 where the object $\ket{\Phi(dA, A)}$  is given by:
 \begin{align}
 \ket{ \Phi(dA, A)} & =  \sum_{i_1, \cdots, i_N}  v_L^\dagger dA^{i_1}_1 A^{i_2}_2 \cdots A^{i_N}_N  v_R \ket{ i_1, i_2, \cdots, i_N } \\ \nonumber
 &  +  \sum_{i_1, i_2, \cdots, i_N}^d  v_L^\dagger A^{i_1}_1 dA^{i_2}_2 \cdots A^{i_N}_N  v_R \ket{ i_1, i_2, \cdots, i_N } \\ \nonumber
 &  +\ \ \ \ \ \ \ \  \cdots \cdots \\ \nonumber
&  +  \sum_{i_1, i_2, \cdots, i_N}^d  v_L^\dagger A^{i_1}_1 A^{i_2}_2 \cdots dA^{i_N}_N  v_R \ket{ i_1, i_2, \cdots, i_N } \\ \nonumber
& := \sum_\alpha dA^{\alpha} \frac{\partial}{\partial A^{\alpha}} \ket{\Psi(A)} \equiv  \sum_\alpha dA^{\alpha}  \ket{ \partial_\alpha \Psi(A)}   \ .
 \end{align}
The $\alpha$ index on the last line combines physical, virtual, and site indices.  Unlike elements in the variational class of MPS, the objects  $\ket{ \Phi(dA, A)}$ form a vector space - and are in fact tangent vectors to the MPS manifold  \cite{2013PhRvB..88g5133H, 2012arXiv1210.7710H}. The gauge invariance of MPS  (\ref{eq:gauge_transf}) can be shown to imply the invariance of tangent states under
\begin{align}
\label{eq:tangent_gauge_inv}
dA^{i_n}  \rightarrow dA^{i_n} + X_{n-1} A^{i_n} - A^{i_n} X_n  \ ,
\end{align}
where the matrices $X$ live in the Lie algebra of the gauge group.

In this paper we will study MPS directly in the thermodynamic limit ($N\rightarrow \infty$), and will use a translation invariant Ansatz, taking the MPS tensors to be position independent (which clearly requires a constant bond dimension $D$).   We refer to the class of such states as uniform MPS (uMPS). Graphically a uMPS is represented as: 
\newline
\centerline{  \includegraphics[scale=0.3]{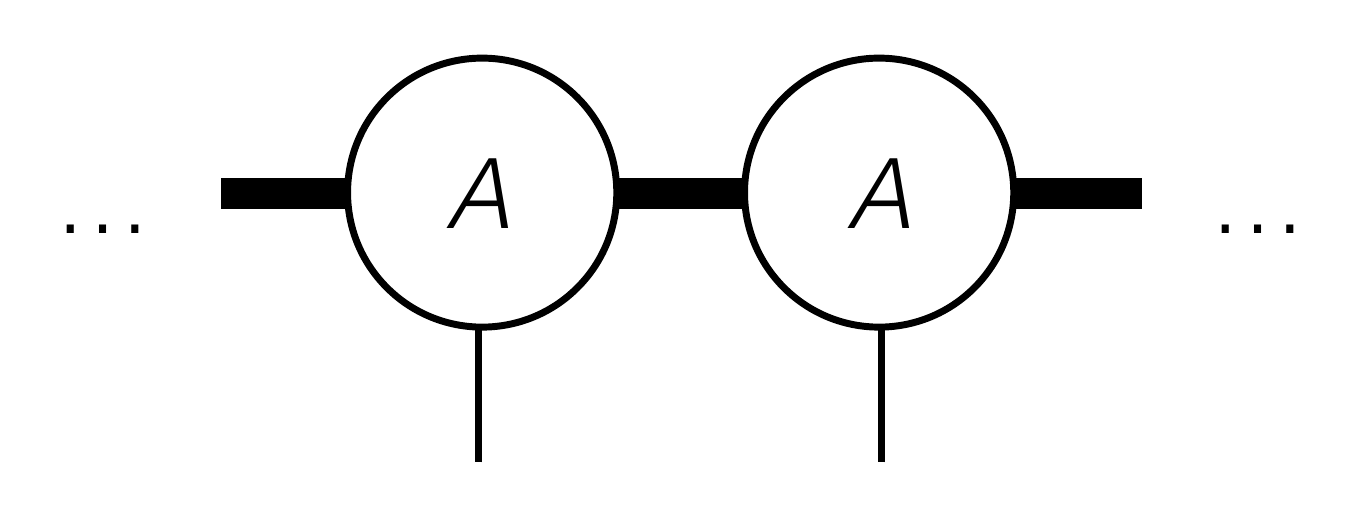} \ \ \raisebox{+2.5ex}{.}  } \newline
The largest eigenvalue of the uMPS transfer matrix $E$,
\begin{align}
\label{uMPS_transfer_matrix}
E := \sum_i A^i \otimes \overline{A}^i   \ , 
\end{align}
 needs to be fixed to unity in order to ensure finite normalisation. The state norm is given by  $( l | r)$, where $(l|$ and $|r)$  are respectively the left and right eigenvectors of $E$ corresponding to eigenvalue 1, and since the eigenvectors can be rescaled freely the state can always be normalised to one.
 
 The expectation value of a local operator acting on $n$ sites can now be written as:
\begin{align}
&  \bra{ \Psi [ A ] }  \hatO \ket{ \Psi [ A ] }  = \\ \nonumber
 &  \sum_{i_1, i_2, \cdots, i_n, \atop j_1, j_2, \cdots, j_n \  }^d  ( l |  O_{i_1 \cdots i_n}^{j_1 \cdots j_n}  \left( A^{i_1} \cdots A^{i_n} \right) \otimes \left( \conj{A}_{j_1} \cdots \conj{A}_{ j_n} \right) |r) \ .
\end{align}
The cost of computing this object, using the optimal sequence of contractions, can be seen to scale as $O(n d D^3)$. 
 
The uMPS parametrisation is invariant under gauge transformations:
\begin{align}
\label{eq:uMPS_gauge_transf}
A^i \rightarrow G A^i G^{-1} \ ,
\end{align}
 and these can be used to set  either $\rho_l = \eye$ or $\rho_r = \eye$ (but in general not both), where $\rho_l$ is the $D^2$ dimensional co-vector $(l|$ reshaped to a $D \times D$ matrix, and similarly for $\rho_r$.  The corresponding gauge transformations in the tangent plane (see (\ref{eq:tangent_gauge_inv})) are given by $dA^{i}  \rightarrow dA^{i} + X A^{i} - A^{i} X$, and this gauge freedom can be used to set either:
\begin{align}
\label{eq:uMPS_l_r_tangent_gauge}
(l  |  \sum_i dA^{i} \otimes \conj{A}^{i} = 0 \ \ \ \ \ \mathrm{or}   \ \ \ \  \sum_i dA^{i} \otimes \conj{A}^{i} | r )  = 0 \ ,
 \end{align}
but again in general not both. These are referred to as, respectively, the left  and right tangent space gauge conditions.

The Gram matrix, given by:
\begin{align}
\label{eq:uMPS_gram_matrix}
G_{\overline{\alpha} \beta} := \braket{\partial_{\overline{\alpha}} \Psi(  A) }{\partial_\beta  \Psi(A)     }  \ , 
\end{align} 
defines a natural metric on the uMPS manifold \cite{2012arXiv1210.7710H}. The TDVP equations can be formally expressed as:
\begin{align}
\label{eq:TDVP_equation}
\sum_{\beta} G_{\overline{\alpha} \beta}  \dot{A^\beta} = & -i \bra{ \partial_{\overline{\alpha}} \Psi(  A)  } \hatH (t) \ket{\Psi  (A(t)) }   \ .
\end{align}
Imposing either the left or right tangent gauge condition (\ref{eq:uMPS_l_r_tangent_gauge}) simplifies the equations significantly, and is in fact necessary to eliminate infinities in (\ref{eq:TDVP_equation}) stemming from transformations along the uMPS state itself. The expression for the overlap of two tangent vectors takes the simple form:
\begin{align}
\label{eq:TDVP_lhs}
\sum_{ \overline{\alpha} , \beta} \overline{A}'^{\overline{\alpha}} G_{\overline{\alpha} \beta}  dA^\beta  =    | \bZ | (l | dA \otimes d\overline{A}' |r ) \ ,
\end{align}
since the left (or right) gauge condition implies that all terms for which the $dA$ and $d \conj{A}'$ tensors are not at the same site are zero. The Gram matrix then takes the simple form: 
\begin{align}
\label{uMPS_local_gram}
G = \rho_l \otimes \rho_r \ \ .
\end{align}
 The nature of the right hand side of (\ref{eq:TDVP_equation}) is elucidated by contracting it with $d\overline{A}'^{\alpha}$:
\begin{align}
\label{eq:TDVP_rhs}
 \sum_{ \overline{\alpha} }  d\overline{A}'^{\overline{\alpha}} & \bra{ \partial_{\overline{\alpha}} \Psi(  A)  }  \hatH (t) \ket{\Psi  (A(t)) } =  \\ \nonumber 
  | \bZ |  & \left(    (l | H^{A A}_{d\conj{A}' \conj{A}} |r) + (l | H^{A A}_{ \conj{A} d \conj{A}'} |r)   \right. \\ \nonumber 
& \left. +   (l | H^{A A}_{\conj{A} \conj{A}} ( \eye -  E )^{PI} (A \otimes d \conj{A}' ) \|r )     \right)   \ .
\end{align}
Here $H^{ A A}_{\conj{A} d\conj{A}'}$, for example, stands for the contraction of the two 'ket' indices of a local term in the Hamiltonian, which is assumed to be translation invariant, with two $A$ uMPS tensors, and the contraction of its 'bra' indices with $\conj{A}$ and $d\conj{A}'$ tensors.   $PI$ indicates a pseudo-inverse on the subspace of $D^2 \times D^2$ matrices defined by the projector $\eye_{D^2 \times D^2} - |r) (l|$. This term stems form a summation over all contributions with the $d\conj{A}'$ tensor to the right of the local Hamiltonian terms (the terms when $d\conj{A}'$ is on the left are zero due to the left tangent gauge condition, which is assumed here). 




\section{Contracting \MakeLowercase{u}CPS with computational cost $O(D^3)$}
\label{app:uCPS_contraction}

Let us first consider the following term appearing on the right hand side of (\ref{eq:uCPS_TDVP_equation}) (i.e. in $-i \bra{ \partial_{\overline{ \tilde{\alpha} \beta }} \Psi(  C(t) )  } \hatH (t) \ket{\Psi  (C(t) ) }$):
 \newline
{ \centerline{ \raisebox{+11.6ex}{$-i$} \includegraphics[scale=0.3]{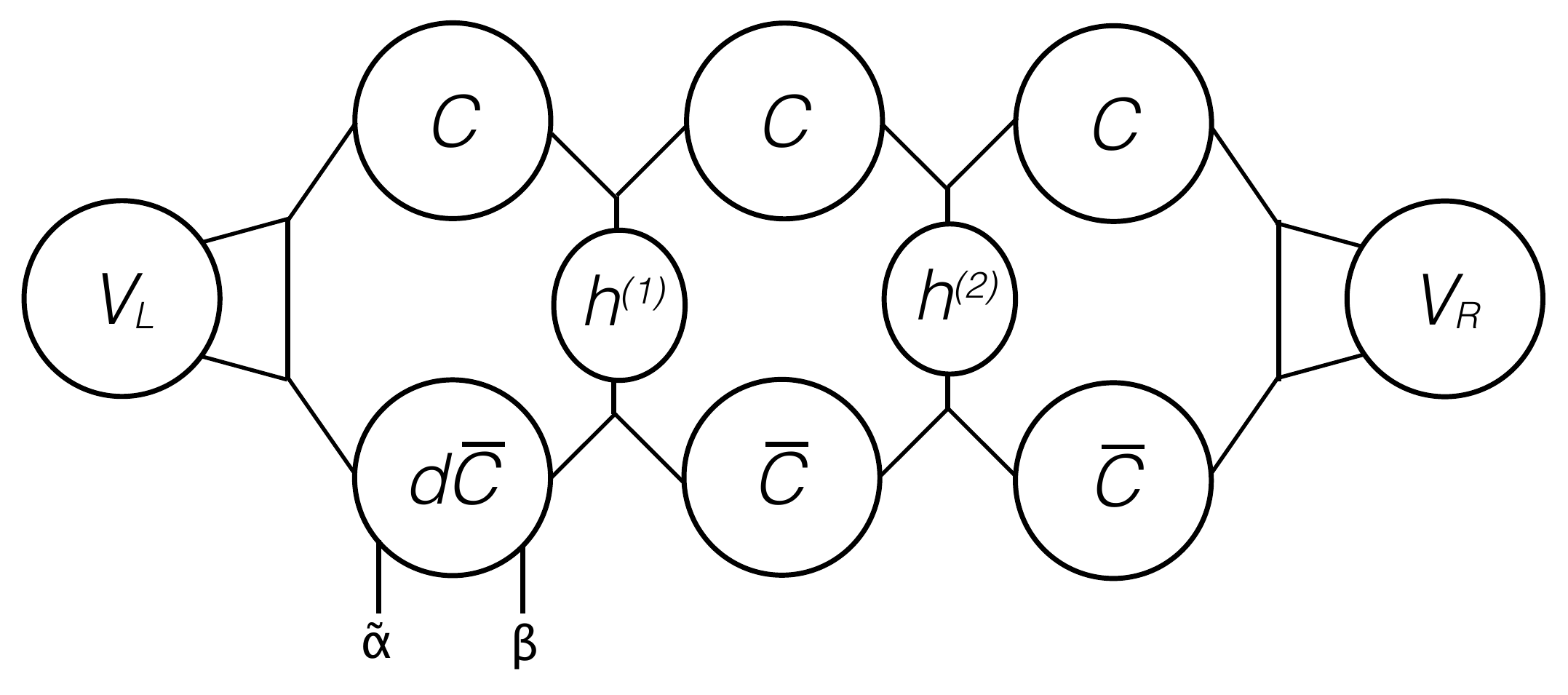}  \ \ \raisebox{+3.5ex}{,} } }  \newline 
where a two-site Hamiltonian has been assumed. Symbolically this corresponds to:
\begin{align}
\label{eq:uMPS_contraction_eg}
 & \left( C_{jm} \right)         \\ \nonumber
T^{\tilde{\alpha} \beta} :=  -i V_{(L) i}  C_{ij}    & \left( h^{(1)}_{jk} h^{(2)}_{mn} \right) C_{m a} V_{(R) a} \ \ , \\ \nonumber
 d \overline{C}^{\tilde{\alpha} \beta}_{ik} &  \left( \overline{C}_{kn} \right)  \overline{C}_{na}
\end{align}
where a summation over all repeated indices is implied, $h^{(1)}$ and $h^{(2)}$ constitute a contribution to a two-site term of the Hamiltonian, and 
\begin{align}
 d \overline{C}^{\tilde{\alpha} \beta}_{ik}  = V_{(R)  i }^{\tilde{\alpha}} V_{(L)  k}^{\beta} (.*) \left( 1 (./) \overline{C}_{ik} \right) \ .
\end{align}
The following contraction ordering (which starts from the right):
\begin{enumerate}
\item  $T^{(1)}_{mn} : = C_{ma} \overline{C}_{na} V_{(R) a}$ is obtained at cost $O(D^3)$,
\item $T^{(2)}_{mn} : = h^{(2)}_{mn} (.*) T^{(1)}_{mn}$ is obtained at cost $O(D^2)$,
\item $T^{(3)}_{jk}  :=  C_{jm} \overline{C}_{kn} T^{(2)}_{mn}$ is obtained at cost $O(D^3)$,
\item $T^{(4)}_{jk} :=  h^{(1)}_{jk} (.*) T^{(3)}_{jk}$ is obtained at cost $O(D^2)$,
\item  $T^{(5)}_{ik} := [ V_{(L)} (.*) C ]_{ij} T^{(4)}_{jk}$ is obtained at cost $O(D^3)$,
\item  $T^{(6)}_{ij} : = T^{(5)}_{ik} (.*) \left (1 (./) \overline{C}_{ik} \right)$ is obtained at cost $O(D^2)$,
\item  $T^{\tilde{\alpha} \beta}  = -iT^{(6)}_{ik} V_{(R)  i }^{\tilde{\alpha}} V_{(L)  k}^{\beta} $ is obtained at cost $O(D^3)$,
\end{enumerate}
can be seen to yield maximal cost $O(D^3)$  at any internal step. It is easy to find other contraction orderings that yield the same cost.

 There are two more contributions to the right hand side of (\ref{eq:uCPS_TDVP_equation}) that are similar to the above.  One of these is  obtained by substituting  $d \overline{C}^{\tilde{\alpha} \beta}_{kn}$ in place of $\overline{C}_{kn}$ in (\ref{eq:uMPS_contraction_eg}), and the other by making the same substitution in place of $\overline{C}_{na}$. A contraction ordering with $O(D^3)$ efficiency can be obtained for these in a very similar manner as what has been demonstrated  above.

The remaining contribution to (\ref{eq:uCPS_TDVP_equation})  corresponds to a sum over all terms with $d \overline{C}^{\tilde{\alpha} \beta}$ not coinciding with the Hamiltonian, and is given by:
 \newline
{ \centerline{  \includegraphics[scale=0.3]{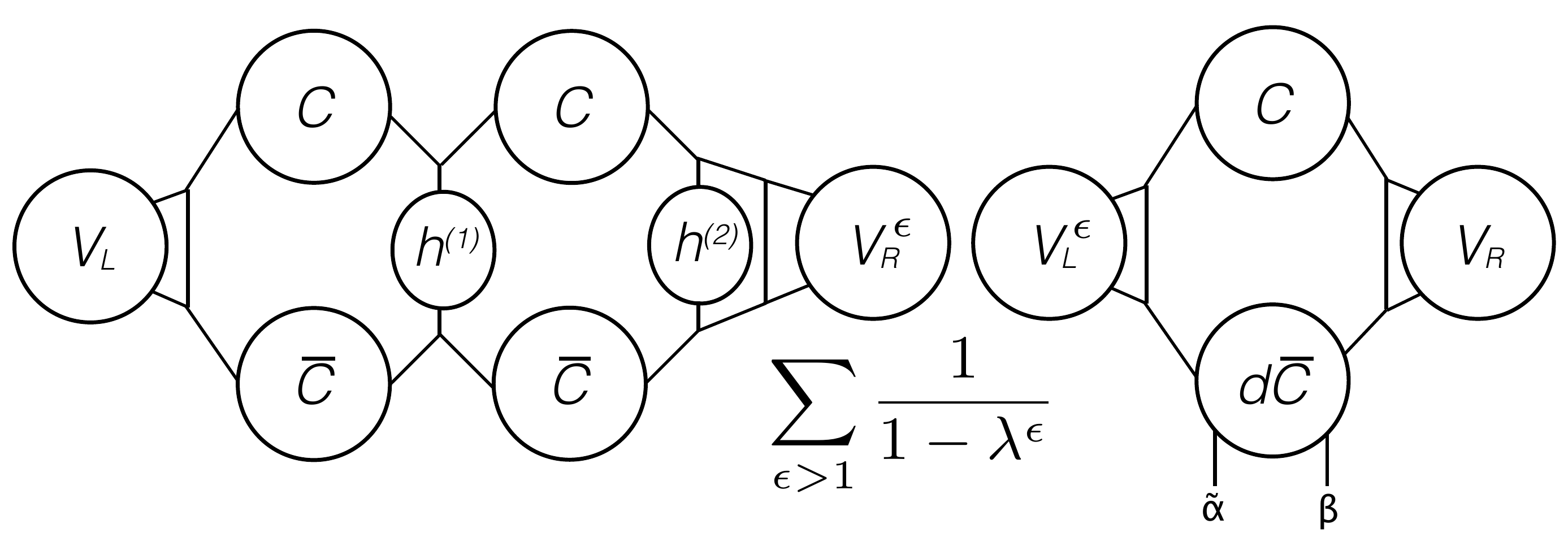}  \ \ \raisebox{+3.5ex}{.} } }  \newline  
The uMPS expression corresponding to this is given by the term on the last line of (\ref{eq:TDVP_rhs}); as with uMPS, the terms with $d\conj{C}$ to the left of the Hamiltonian are zero due to the left tangent gauge condition, which has been assumed here.  It is worth noting that while the computation of the pseudo-inverse acting on a vector in the uMPS expression can only be achieved at cost $O(d D^3)$, by recourse to an iterative procedure such as the biconjugate gradient algorithm, the uCPS term above can be computed explicitly at cost $O(D^3)$ using a contraction ordering similar to that described for (\ref{eq:uMPS_contraction_eg}).

\section{\MakeLowercase{u}CPS Gram matrix inverse pre-conditioning}
\label{app:uCPS_precond}

The uCPS Gram matrix is given by:
\begin{align}
\label{eq:uCPS_gram_matrix_explicit2}
G^{ \tilde{\gamma} \delta | \tilde{\alpha} \beta} = \sum_{ij} \lambda_{ij} T^{ \tilde{\gamma} \delta | \tilde{\alpha} \beta}_{ i j} \ ,
\end{align}
where
\begin{align}
T^{ \tilde{\gamma} \delta | \tilde{\alpha} \beta}_{ i j} = \left(  \conj{V}_{(R) i }^{\tilde{\gamma}} \conj{V}_{(L) j}^{\delta}   \right) (.*) \left(   V_{(R) i }^{\tilde{\alpha}} V_{(L) j }^{\beta}   \right) \ .
 \end{align}
The convention is that the non-tilde indices run over the whole range, $1$ to $D$, while the tilde-indices correspond to a truncation and run from $2$ to $D$, and $V_{(L)}^1 \equiv V_{(L)}$ and $V_{(R)}^1 \equiv V_{(R)}$ are the left and right eigenvectors corresponding to eigenvalue one. 

Let us first consider the matrix:
\begin{align}
\label{eq:inverse_guess}
\tilde{G}^{ \tilde{\gamma} \delta | \tilde{\alpha} \beta} = \sum_{ij} (1 (./) \lambda_{ij} ) \tilde{T}^{ \tilde{\gamma} \delta | \tilde{\alpha} \beta}_{ i j} \ ,
\end{align}
where
\begin{align}
\tilde{T}^{ \tilde{\gamma} \delta | \tilde{\alpha} \beta}_{ i j} = \left(  \conj{V}_{(L) i }^{\tilde{\gamma}} \conj{V}_{(R) j}^{\delta}   \right) (.*) \left(   V_{(L) i }^{\tilde{\alpha}} V_{(R) j }^{\beta}   \right) \ .
 \end{align}
The obstruction to $\tilde{G}$ being the inverse of $G$ can be understood as originating in the truncation of the eigenvalue one eigenvector in our implementation of the left tangent gauge condition (\ref{eq:dC_uCPS_TDVP}). Namely, the eigenvector matrices obey:
\begin{align}
 V_{(L) i}^{\delta} V_{(R) i}^{\epsilon} & = \delta^{ \delta \epsilon} \ , \\ \nonumber
 \overline{V}_{(R) j}^{\epsilon} V_{(L) m}^{\epsilon} & = \delta_{jm} \ ,  \\ \nonumber
 V_{(R) i}^{\tilde{\delta}} V_{(L) i}^{\tilde{\epsilon}} & = \delta^{\tilde{\delta} \tilde{\epsilon}}   \ , \\ \nonumber
 \overline{V}_{(L)_j}^{\epsilon} V_{(R) m}^{\epsilon} & = \delta_{jm} \ , \\ \nonumber
 V_{(L)j}^{\tilde{\gamma}} \overline{V}_{(R) m}^{\tilde{\gamma}} & = \delta_{jm} - V_{(L) j } V_{(R) m } \ , \\ \nonumber
 \overline{V}_{(R) j}^{\tilde{\gamma}} V_{(L)m}^{\tilde{\gamma}} & = \delta_{jm} - V_{(R) j} \overline{V}_{(L)m} \ ,
\end{align}
where summation over repeated indices is understood. The presence of projectors in the last two lines expresses the deformation of the exact orthogonality relations due to the $\alpha \rightarrow \tilde{\alpha}$ truncation, which prevents (\ref{eq:inverse_guess}) from being the Gram matrix inverse.

In order to explicitly compute the action of the inverse of $G$ on a vector with computational cost scaling as $O(D^3)$, it is necessary to be able to express $G^{-1}$ in the general form (\ref{eq:inverse_guess}), or at least as a sum of a constant number of terms of this form. We have not managed to find any suitable solution that would bypass the obstruction described above (but have also not proved that doing so is impossible).  However, in order to achieve $O(D^3)$ scaling, one can instead make use of an iterative algorithm, the biconjugate gradient (stabilised) method. This algorithm provides a solution for $\vec{x}$ in the equation $\mathbb{A} \vec{x} = \vec{b}$, where $\mathbb{A}$ is some invertible matrix; as input only the action of the matrix $\mathbb{A}$ on a vector needs to be supplied. For the present problem this  can be achieved with cost $O(D^3)$, using a contraction scheme along the lines of Appendix \ref{app:uCPS_contraction}. 

Achieving $O(D^3)$ scaling assumes that, in the limit of large bond dimension $D$, the number of iterations needed for the biconjugate gradient algorithm to converge to some desired accuracy scales as roughly a constant. In practice this may not always be the case. Using the biconjugate algorithm as described above,  $O(D^3)$ scaling is indeed spoiled for a general uCPS TDVP computation. This is ultimately related to the fact that uCPS, viewed as a restriction of uMPS, fixes nearly all the gauge freedom, which in general causes $G$ to become badly conditioned in the course of a TDVP run -  even when Schmidt values of the uCPS state itself are much larger than machine precision. One solution to this problem is to make a judicious choice of a pre-conditioner matrix. In general, this refers to a  matrix $\mathbb{M}^{-1}$, in $\mathbb{M}^{-1} \mathbb{A} \vec{x} =\mathbb{M}^{-1} \vec{b}$, which can be employed in the iterative algorithm in order to make the problem better conditioned.  For the present case, taking $\mathbb{M}^{-1} = \tilde{G}$ in fact seems to be the optimal choice (again, the iterative algorithm only needs to be supplied with the action of $\tilde{G}$ on a vector, which can be achieved with cost $O(D^3)$). It should be noted that without the pre-conditioning step, for most of the examples in this paper,  it would only have been practically feasible to compute quenches and ground state approximations with very small overlap sizes, and many of the computations performed in this paper would not have been accesible.

\end{small}

\newpage

\bibliography{bibliography}

\end{document}